\documentclass[9pt,twocolumn,twoside]{osajnl}
\usepackage{subcaption}
\usepackage{comment}
\usepackage{soul}
\usepackage{xcolor}
\usepackage{booktabs}
\journal{josab} % Choose journal (ao, aop, josaa, josab, ol, optica, pr)

\setboolean{shortarticle}{true}
\usepackage{notes2bib}
\usepackage{ulem}
\usepackage{makeidx}

\newcommand{\ket}[1]{| #1 \rangle}
\newcommand{\bra}[1]{\langle #1 |}

\newcommand*{\thead}[1]{\multicolumn{1}{c|}{\bfseries #1}}

\title{Quantum imaging and metrology with undetected photons: a tutorial}

\author[1,*]{Gabriela Barreto Lemos}
\author[2]{Mayukh Lahiri}
\author[3,4]{Sven Ramelow}
\author[5]{Radek Lapkiewicz}
\author[6]{William N. Plick}
\affil[1]{Instituto de F\'isica, Universidade Federal do Rio de Janeiro, Av. Athos da Silveira Ramos 149, Rio de Janeiro, CP: 68528, Brazil.}
\affil[2]{Department of Physics, Oklahoma State University, 145 Physical Sciences Bldg, Stillwater, Oklahoma 74078, USA}
\affil[3]{Humboldt-Universität zu Berlin, Institut für Physik, Newtonstraße 15, 12489 Berlin, Germany}
\affil[4]{IRIS Adlershof, Humboldt-Universität zu Berlin, Zum Großen Windkanal 6, 12489 Berlin}
\affil[5]{Institute of Experimental Physics, Faculty of Physics,
University of Warsaw, Pasteura 5, Warsaw 02-093, Poland}
\affil[6]{University of Dayton, Department of Physics, Department of Electro-Optics, Dayton, OH, 45469, United States}
\affil[*]{Corresponding author: gabrielabl@if.ufrj.br}

 \dates{Compiled \today}

\begin{abstract}
We present a tutorial on the phenomenon of induced coherence without induced emission, and specifically its application to imaging and metrology. It is based on a striking effect where two nonlinear crystals, by sharing a coherent pump and one of two output beams each, can induce correlations between the other two individual, non-interacting beams. This can be thought of as a type of quantum-erasure effect, where the ``welcher-weg'' (which-way), or in this case ``which-source'' information is erased when the shared beams are aligned. With the correct geometry this effect can allow an object to be imaged using only photons which have never interacted with the object \--- in other words the image is formed using undetected photons. Interest in this and related setups has been accelerating in recent years due to a number of desirable properties, mostly centered around the fact that the fields for detection and imaging (since separate) may have different optical properties, entailing significant advantages to various applications. The purpose of this tutorial is to introduce researchers to this area of research, to provide practical tools for setting up experiments as well as understanding the underlying theory, and to also provide a comprehensive overview of the sub-field as a whole.    
\end{abstract}

\setboolean{displaycopyright}{true}

\begin{document}

\maketitle
\section{Introduction}

One thousand years ago, in 1021, \d{H}asan Ibn al-Haytham (\textit{a.k.a.} Alhazen) completed the ``Book of Optics'' in which he laid the foundations of modern optics and detailed the apparatus that later Kepler dubbed \textit{camera obscura} (pinhole camera). In the intervening ten centuries imaging technology has progressed immeasurably \--- and up until the present day, developing ever-better imaging and sensing devices remains an extremely active field of research. Evolving approaches to imaging technology have enabled new abilities like extreme sensitivity and resolutions, imaging at non-visible wavelengths, and many others too diverse to fully enumerate.
\par
One of the very-latest areas of research is a suite of new technologies enabled by the development of quantum theory, which contains the most accurate description of optical fields. Devices that take advantage of the co-called ``quantum nature of light'' include new imaging schemes which may beat the classical limits of sensitivity \cite{brida2010experimental,sabines2019twin,garces2020quantum,Casacio2021}, spatial resolution \cite{santos2003resolution, santos2005generation, schwartz2013superresolution,classen2017superresolution,unternahrer2018super,tenne2019super}, and phase metrology \cite{giovannetti2011advances, polino2020photonic, toth2014quantum}. Quantum Optics has also enabled the emergence of imaging schemes where the light that interacts with the sample is \emph{not} captured by the pixelated detector/camera, \textit{e.g.} Interaction Free Imaging \cite{white1998interaction}, Ghost Imaging  \cite{klyshko1988effect,belinskii1994two,pittman1995optical,gatti2008quantumimaging,chan2009two,aspden2013epr, moreau2019imaging}, and \textit{Quantum Imaging with Undetected Photons} (QIUP)  \cite{Lemos:2014p346,   Lahiri2015a, viswanathan2021position}. 
\par
With ever-more-sophisticated camera and photon-source technologies emerging in recent years, quantum imaging and quantum-inspired imaging techniques have become even more promising avenues for research and development \--- and will likely be a key component of the twenty-first century quantum revolution, alongside quantum computing, quantum communication and quantum metrology. 
\par
Here we focus on \textit{Quantum Imaging with Undetected Photons} (QIUP) and its spin-offs. In these experiments, coherence is induced between light produced in two twin-photon sources placed within a nonlinear interferometer. Quantum imaging with nonlinear interferometers was introduced in \cite{Lemos:2014p346} and applications of this method to bio imaging, spectroscopy, optical coherence tomography and moving images were first proposed in \cite{patent1,patent2}. The most important advantage of this kind of technique is that one can obtain information about an object probed by a light beam of one wavelength by only detecting a \emph{separate} light field at a different wavelength. The light field which illuminates the sample is not detected at all. This is especially useful when the illumination wavelength is one for which detectors are not available or unsatisfactory, and in the case of delicate samples that require low intensity illumination.
\par
\textbf{How to read this tutorial:} In this tutorial we will prepare you both theoretically and experimentally to investigate a few methods of quantum imaging and metrology with undetected photons, i.e., quantum imaging based technologies using \textit{induced coherence without induced emission} within a nonlinear interferometer in the low gain regime. We will assume basic knowledge of the quantum optics formalism and familiarity with spontaneous parametric down-conversion (SPDC). This tutorial does not have to be read in a linear fashion. Readers with different interests might jump to different sections. In particular, those interested in theory may skip sections \ref{sec:design}, \ref{sec:tips}. Those who are not planning on doing detailed calculations, but are interested in building an experiment for imaging or metrology with undetected photons may skip sections \ref{sec:operators}--\ref{sec:rig-theory}. 
\par
If this is your first encounter with imaging and metrology using induced coherence without induced emission (ICWIE), we suggest you start by reading Sec.~\ref{inducedcoherence}. In Sec.~\ref{sec:operators} we give a basic overview of the main quantum optics states and operators, for those who have not encountered quantum optics formalism before or would like a recap. Readers interested in the application of ICWIE to phase metrology can learn about it in Sec.~\ref{sec:metrology}. In Sec.~\ref{sec:rig-theory} we will provide the  theoretical description for a multi-mode nonlinear interferometer and its application to phase and absorption imaging. In Sec.~\ref{sec:applications} we describe optical coherence tomography, holography and spectral imaging with undetected photons. The following sections, \ref{sec:design} and \ref{sec:tips}, are aimed at giving experimental guidelines to researchers who are building a Zou-Wang-Mandel (ZWM) interferometer or a SU(1,1) interferometer in the low gain regime. 
In Sec.~\ref{outlook}, we give an outlook of some interesting research directions one could explore.
\par
 Before diving into the tutorial, we would like to point out that in our exposition we use the language of squeezed correlated fields, such as states produced in Parametric Down Conversion. However, the interferometers we analyse and the mathematical formalism we introduce can also be applied to a variety of other systems, such as atomic spin waves \cite{Chen2015}, superconducting microwave cavities \cite{lahteenmaki2016, bruschi2017}, and four wave mixing \cite{ou2017quantum}.
 \par
\section{Induced Coherence without Induced Emission}\label{inducedcoherence}
\par
Richard Feynman considered quantum interference the biggest mystery in quantum mechanics \cite{feynmanIII}. Quantum interference is a phenomenon observed at a detector if and only if it is impossible to associate each detected quantum (\textit{e.g.} photon, electron, atom, molecule) to a particular path, among the 
two or more paths connecting that detection apparatus to the quantum source \cite{feynmanIII, mandel1991coherence}.
Moreover, the fringe visibility in a two-way interferometer gives an upper bound on the available which-way (\textit{welcher weg}) information \cite{englert1996fringe}. The quintessential example is the double slit experiment, where no interference is observed if the alternative paths between the source and the detector are distinguishable and interference is observed if those  paths are indistinguishable \cite{feynmanIII, Sudarshandouble}. Other important examples in optics are the Michelson,  Mach-Zehnder and Sagnac interferometers \cite{Zetie_2000,rauch2015neutron, bachor2004guide}. 
\par
\begin{figure}
    \centering
   \includegraphics[width=1\linewidth]{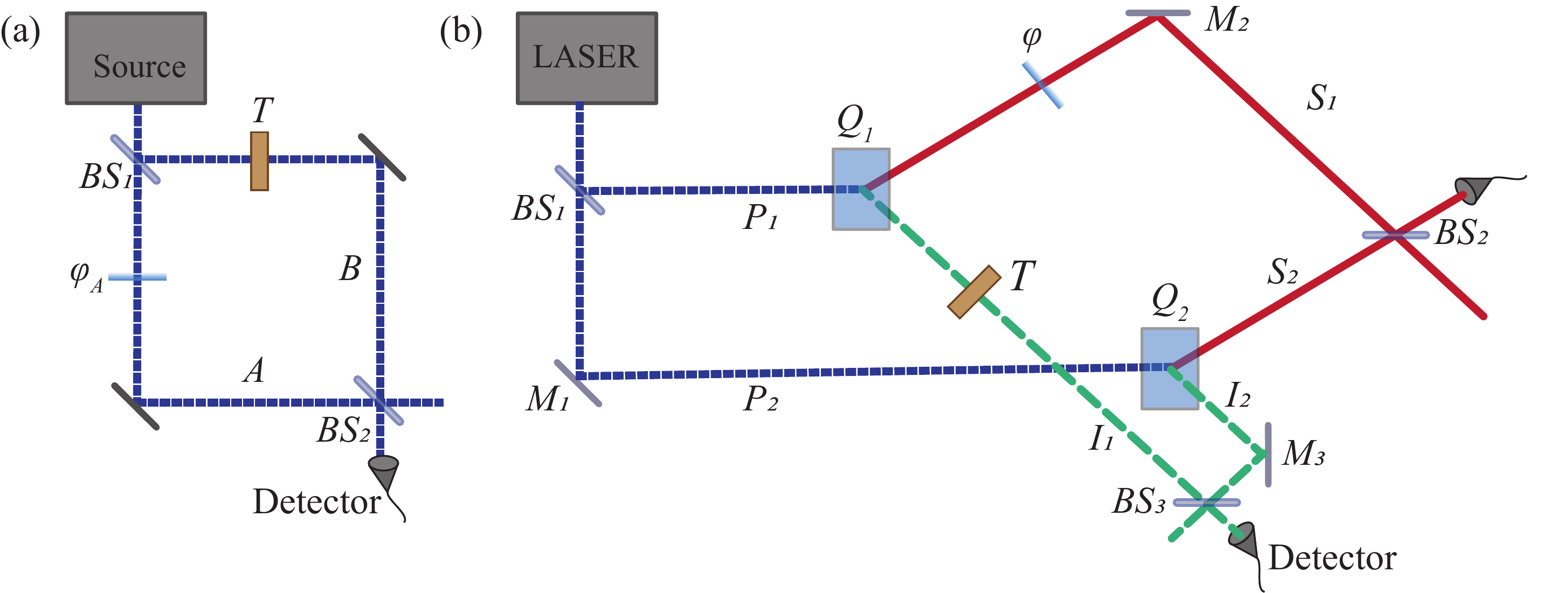}
    \caption{(a) \textbf{Mach-Zehnder interferometer}. A light field is split at the $50$:$50$ beam splitter, BS$_1$, and is recombined at BS$_2$. A phase shifter is placed in path $A$ and an object with complex field transmittance $T$ placed in path $B$. Interference is analysed at the detector. (b) \textbf{Two-Particle Interferometer}. A pump is split into paths $P_1$ and $P_2$ at a $50$:$50$ beam splitter (BS$_1$) and illuminates two nonlinear sources, $Q_1$ and $Q_2$, producing correlated particle pairs. When a pair is produced in source $Q_1$ ($Q_2$), the so-called {\it signal} particle is emitted into path $S_1$ ($S_2$) and so-called {\it idler} particle is emitted into path $I_1$ ($I_2$). Signal paths $S_1$ and $S_2$ are combined at BS$_2$ and idler paths $I_1$ and $I_2$  are combined at BS$_3$. No interference is observed in the signal intensity at the detector because which way information is {\textit in principle} obtainable. One can only observe interference using post-selection, {\textit i.e.}\ by detecting idlers at one output of BS$_3$ \textit{in coincidence} with signals at one output of BS$_2$. }
    \label{fig:Mach-Zehnder}
\end{figure}
\par
When setting up any interferometer in the laboratory, in order to observe good  interference visibility, one aligns the beams incoming to the detector and adjusts the path lengths of the interferometer so that they differ by not more than the coherence length of the quanta. In the language of quantum information, the alignment and the length adjustment of the interferometer paths amount to ensuring indistinguishability between quanta arriving at the detector, thus enabling interference \cite{mandel1991coherence}. 
\par
 \textbf{The Mach Zehnder Interferometer}. Let us consider what happens to light in a Mach-Zehnder interferometer (MZI), illustrated in Fig.\ref{fig:Mach-Zehnder}. The incoming light field is split at the first beam splitter BS$_1$ and recombined at a second beam splitter, BS$_2$. For interference to be seen the optical path lengths of paths $A$ and $B$ between the two beam splitters must be equal to within the coherence length of the light field, such that such that each photon is described as being in a superposition, $\left(\ket{1}_A+e^{-i{\phi_A}}\ket{1}_B\right)/\sqrt{2}$,  where $\ket{1}_x$ denotes a photon in path $x$. The phase $\phi_A$ is associated with relative optical delays between the two modes of the interferometer and can be adjusted by slightly shifting a mirror or beam splitter or by inserting a slab of a transparent material, such as silica. 
 The count rate at a detector placed after BS$_2$ in path $A$ (or $B$)  is given by 
 \begin{equation}
    \mathcal R_{A/B}=\frac{\left(1+|T|^2\right)\pm 2|T|\cos(\phi-\gamma)}{4},  
 \end{equation}
 where $T=|T|e^{i\gamma}$, with $0\leq |T|\leq 1$ and $0\leq \gamma<2\pi$,  is the complex field field transmittance of an object placed in path $B$. The interference visibility obtained by scanning $\phi_A$, is defined as 
 \begin{equation}\label{vis-def}
\mathcal V=\frac{\mathcal{R}_{\max}-\mathcal{R}_{\min}}{\mathcal{R}_{\max}+\mathcal{R}_{\min}}.
 \end{equation}
Assuming equal optical path lengths, the interference visibility of the MZI is thus 
\begin{equation}
    \mathcal{V}_{MZ}=\frac{2|T|}{1+|T|^2}.
    \label{eq-MZ-vis}
\end{equation} 
In the language of quantum information, the reduction of visibility for $|T|<1$ is due to the path distinguishability introduced by the object. 
One application of the MZI is \textit{interaction free measurements} \cite{elitzur1993quantum,kwiat1995interaction}, which can be used for imaging \cite{white1998interaction}. An object is placed in one arm of a MZI where photons are sent one at a time. 
The object affects the interference pattern at the output and,  in a fraction of the experimental runs, the presence of the object can be deduced without the photon having interacted with it. 
\par
\textbf{The Zou-Wang-Mandel (ZWM) Interferometer}.
 In the MZI, as well as the Michelson Interferometer and the Sagnac,  classical wave models, including classical electromagnetism, can describe the interference visibility due to (mis)alignment, path length difference and/or an object placed in a path of the interferometer. Seeking to unravel the connection between quantum indistinguishability and interference,  in $1991$ Zou, Wang and Mandel, with an essential insight from Jeff Ou, created an interferometer which cannot be described by classical wave optics models \cite{zou1991,Wang:1991p625}.
\par
\begin{figure*}[t]
\centering
\includegraphics[width=\linewidth]{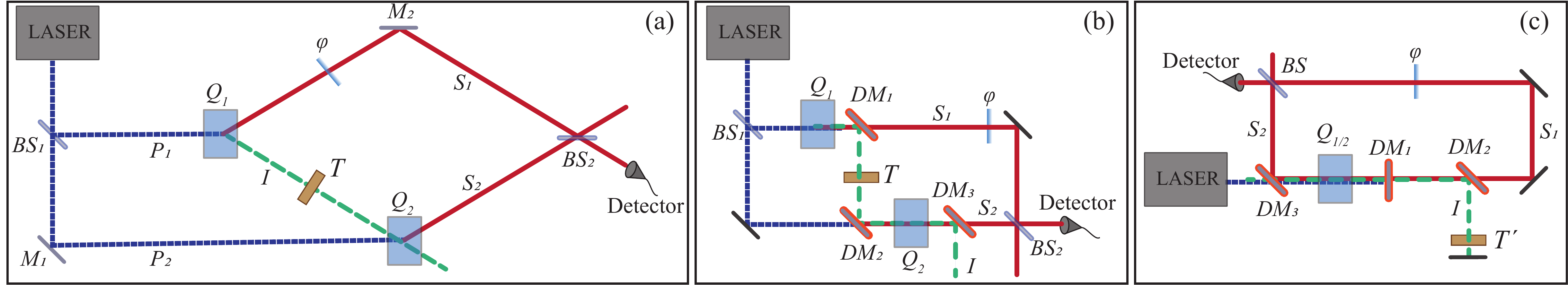}
\caption{\textbf{Three architectures of the Zou-Wang-Mandel Interferometer}. In (a) and (b) the idler paths produced in sources $Q_1$ and $Q_2$ are aligned and a $50$:$50$ beam splitter, BS$_2$, combines signal paths $S_1$ and $S_2$, {\it which path} information is erased and single-particle interference can be observed in the detector, even though idler particles are not detected at all. The field transmittance function, $T$, of an object illuminated by the idler field $I$ can be observed in the interference pattern of the signal beams at the detector, even though signal particles have not interacted with that object. The phase $\phi$ can be tuned by adjusting signal, idler or pump optical path lengths. In (b) and (c) signal and idler are emitted in the same direction as the pump (collinear emission), and if they are all at distinct frequencies, dichroic mirrors can be used to separate them. In (b) the dichroic mirror DM$_1$ reflects idler photons and transmits signal wavelength, whereas DM$_2$ transmits the pump and reflects idler photons. In (c), single crystal is pumped from both sides. Dichroic mirror, DM$_1$, reflects the pump and transmits both signal and idler, whereas DM$_2$ reflects idler and transmits the signal. Finally, DM$_4$ reflects signal and idler while transmitting the pump. Notice that in this architecture, undetected light goes twice through the same sample ($T'=T^2$).}
\label{fig:basicmandel}
\end{figure*}

 
A ZWM setup uses two identical sources of photon pairs, e.g. nonlinear crystals that can generate photon pairs through spontaneous parametric down-conversion (SPDC). The sources are prepared such that the biphoton fields emerging from them are mutually coherent. These crystals are weakly pumped by mutually coherent laser beams, for example, generated by splitting a laser beam into two as shown in Fig.~\ref{fig:basicmandel}. Let us denote the two sources $Q_1$ and $Q_2$ and the emitted beams will be referred to as signal beams ($S_1$ and $S_2$) and idler beams ($I_1$ and $I_2$). The beams $S_1$ and $S_2$ are combined at a beam splitter, the outputs of which are sent to detectors. Both sources emit the idler beams into the same spatial mode, $I$. Considering that the  the idler photons are not detected at all, do you expect that interference fringes can be observed in the detected signal outputs? Why or why not?
\par 
\par
A few simple calculations can help us understand what is going on.  In a first approach to this problem, it is instructive to write down the particle states (here photons) in the device as relatively-simple state vectors. In this picture the action of a nonlinear source is simply to add one photon each to the appropriate modes. For example, the action of first source, $Q_1$ in Fig.~\ref{fig:basicmandel} is modelled as taking the vacuum input in modes $S_1$ and $I_1$ and transforming it as $|0\rangle_{S_1} |0\rangle_{I} \rightarrow |1\rangle_{S_1} |1\rangle_{I}$, where $|1\rangle_{S_1}, \;|1\rangle_{I}$ represent a photon occupying mode $S_1, \;I$, respectively. We keep the kets as single-photon number states so as to maintain the same notation as the rest of the paper, but since first quantization does \emph{not} use the Fock basis (all modes are assumed to have only one photon) we could equally-well omit the occupation number \--- that is $|1\rangle_{I}=|I\rangle$, for example.
Given this assumption it is also very unlikely that \emph{both} nonlinear sources, $Q_1$ and $Q_2$, fire in sync.  We will see in Sec.~\ref{sec:rig-theory} that the object/sample in the idler path is modelled as a beam splitter with complex field field transmittance function of that object by $T=|T|e^{i\gamma}$, with $0\leq|T|\leq 1$ and $0\leq\gamma<\pi/2$. In this picture, the action of the object in Fig.~\ref{fig:basicmandel} is to transform the state produced in source $Q_1$ as $|1\rangle_{S_1} |1\rangle_{I} \rightarrow 
 |1\rangle_{S_1} 
\left( T|1\rangle_{I}+i\sqrt{1-|T|^2}|1\rangle_0\right) $, where $|1\rangle_0$ represents a photon absorbed or scattered by the object.
\par
Assuming both sources are identical and their emissions are coherent, the state of a photon pair just \textit{before} BS$_2$ in Fig.~\ref{fig:basicmandel} can be written as 
\begin{equation}
 \frac{ 
\left( T|1\rangle_{I}+i\sqrt{1-|T|^2}|1\rangle_0\right)|1\rangle_{S_1} +e^{-i\phi}|1\rangle_{I} |1\rangle_{S_2}}{\sqrt{2}},
\end{equation}
where $0\leq\phi<2\pi$. Note that mode $I$ does not acquire a second photon as the single photon in that mode is assumed to have come from \emph{either} the second or first crystal. At this point the origin of this photon could be determined by seeing which of modes $S_1$ and $S_2$ contain a photon. 
\par
The final beam splitter BS$_2$ combines the signal fields, after which the state of the twin particles can be written as $|\psi_f\rangle=$
\begin{align}
|\psi_f\rangle=&\frac{1}{2}\left( \left(T+ie^{-i\phi}\right) |1\rangle_I +i \sqrt{1-|T|^2}|1\rangle_0 \right) |1\rangle_{S_1}\nonumber \\
&+  \frac{1}{2}\left( \left(e^{-i\phi}+i T\right) |1\rangle_I +\sqrt{1-|T|^2}|1\rangle_0 \right) |1\rangle_{S_2}%\nonumber
 \end{align} 
By tracing out the idler modes $I$ and $O$ in the state above, we can obtain the count-rate at a detector placed at either output of BS$_2$:
 \begin{eqnarray}\label{eq:zwm-counts}
 \mathcal{R}_{S_1(S_2)}=\frac{1\pm|T|\cos(\phi+\gamma)}{2}.
 \end{eqnarray}
 \par
\noindent Where \--- strikingly \--- an interference pattern modulated by the object $T$ can now be observed, despite the fact that \emph{neither} $S_1$ or $S_2$ interacted with that object. Coherence is thus ``induced'' between the two modes as the result of aligning as precisely as possible the shared idler mode. The idler mode $I$, which contains no phase information, is typically discarded. 
\par
 In the ZWMI, which we have just described, the interference visibility is directly proportional to the absolute value of the field transmission coefficient:
 \begin{equation}\label{vis-zwm}
     \mathcal{V}_{ZWM}\propto |T|.
 \end{equation} 
This relationship holds even if the intensities of the signal beams are not equal. In fact, any photon loss in the idler arm results in a reduction of the total visibility, as it introduces partial path distinguishability (\textit{welcher-weg} information).
 \par
Let us compare this with the nonlinear effect of loss ($|T|<1$) in an arm of a MZI (Eq.\ref{eq-MZ-vis}). The linear relation between interference visibility and loss in the undetected idler photon path between the two sources is a distinguishing feature of the ZWMI interferometer. It is shown in \cite{zou1991,wiseman2000} that this characterizes the non-classicality of induced coherence \textit{without induced (stimulated) emission}. This is a very important point: \textit{stimulated emission at the second source $Q_2$ due to the input idler field $I$ is not necessary for induced coherence (interference)}, a fact which highlights the non-classicality of the phenomenon \cite{lahiri2019}. In the cases where stimulated emission is not negligible in $Q_2$, the interference visibility is a nonlinear function of the field field transmittance $|T|$ \cite{wiseman2000,kolobov2017}. This regime can be achieved using very high gain sources (\textit{e.g} using very high pump power), or by seeding $Q_1$ and $Q_2$ via mode $I$ with a coherent state (a laser beam) with the idler beam wavelength \cite{ou1990coherence,wang1991observation}. 
 \par
 The ZWM interferometer was generalized to many spatial modes using spatially correlated photon pairs to produced {\it Quantum imaging with undetected photons} (QIUP) \cite {Lemos:2014p346,  Lahiri2015a}. Quantum  interference and spatial correlations between signal and idler photons \cite{walborn10} together produce in the detected signal photons images of an  object placed in the idler beam (path $d$) \cite{Lahiri2015a, Armin2016a, Lahiri2016, viswanathan2021position}. Both the real part and the imaginary part of the object's spatially varying field transmittance function can be observed. The real part $|T|$ is encoded in the interference visibility. The imaginary part $\gamma$ appears as an interferometric phase.
 \par
 \textbf{Two-Particle Interferometer}.
A natural question to ask is what happens if, instead of both sources emitting into the same spatial idler beam, $I$ (Fig.~\ref{fig:basicmandel}), the idler fields from $Q_1$ and $Q_2$ were emitted into separate spatial modes, $I_1$ and $I_2$, and only later combined at a beamsplitter ($BS_3$), as illustrated in Fig.~\ref{fig:Mach-Zehnder}(b).  A straightforward calculation shows that in this case, interference is \textit{not} directly observed in intensity measurements at the signal detector without any post-selection (coincidence detection). 
The two photon state after BS$_2$ and BS$_3$ in Fig.~\ref{fig:Mach-Zehnder}(b) is
\begin{align}
\left(\frac{f_-|1\rangle_{I_1}+i f_+ |1\rangle_{I_2}  +i \sqrt{2(1-|T|^2)}|1\rangle_{0}}{2\sqrt{2}}\right)| 1\rangle_{S_1}\nonumber \\ +\left(\frac{i f_+|1\rangle_{I_1}-f_- |1\rangle_{I_2}  -\sqrt{2(1-|T|^2)}|1\rangle_{0}}{2\sqrt{2}}\right)| 1\rangle_{S_2},
 \end{align}
 where $f_\pm\equiv\left(Te^{i\phi}\pm1\right)$.
 
 If the idler photons remain undetected, mathematically we perform a partial trace over the idler modes and obtain a constant (=$1/2$) signal photon counting rate at either output of BS$_2$. In other words, no interference is observed in signal intensities. An intuitive explanation for this is that which-source information is retrievable \textit{in principle}, for example, one could (hypothetically) add a fourth Beam Splitter combining the two idler outputs, which would reveal which idler came from each source. In other words, by not detecting idlers we are leaving "lose ends" of information which inhibit signal intensity modulation due to interference.
 \par
 Note that if one detects one output of BS$_2$ \textit{in coincidence with} an output of BS$_3$, \textit{i.e.} by using post-selection, it is possible to observe interference \cite{ou1990coherence, horne1989two}, and the visibility is  $\mathcal{V}=2|T|/(|T|^2+1)$.
 \par
\begin{figure*}[t]
\centering
\includegraphics[width=\linewidth]{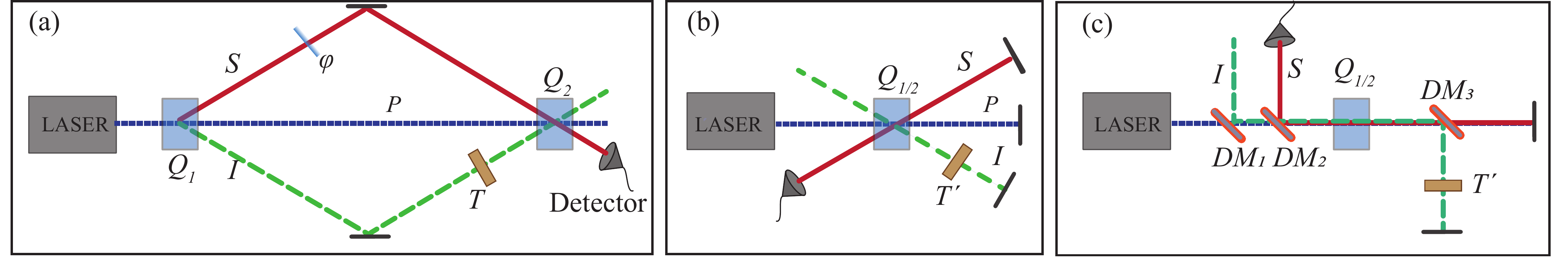}
\caption{\textbf{Three architectures for the SU(1,1) Interferometer.} In (a) both signal and idler paths $S$ and $I$ from the first nonlinear source ($Q_1$) are aligned with signal and idler paths originating in the second nonlinear source ($Q_2$). In (b) and (c) the laser pumps a crystal and is reflected back through the same crystal The undetected light traverses twice through the same sample ($T'=T^2$). In (c) the pump, signal and idler leave the crystal collinear with each other and if they are all at different frequencies they can be separated using three dichroic mirrors: DM$_1$ transmits the pump but reflects signal and idler photons; DM$_1$ transmits the pump and idler photons, but reflects signal photons; DM$_3$ transmits the pump and signal photons, but reflects idler photons. In all three architectures, single-particle interference is seen in signal and idler outputs \textit{without} post-selection (coincidence detection). The field transmittance function of an object placed in either paths $S$ or $I$ can be seen in the interference pattern appearing in a camera placed in either output path. }
\label{fig:herzog}
\end{figure*}
\par
\textbf{An SU(1,1) Interferometer}. Let's now consider that both signal and idler photons from source $Q_1$ are fed into source $Q_2$ \cite{yurke19862, Herzog}, as shown in Fig.\ref{fig:herzog}. Here we will refer to this interferometer as an ``SU(1,1) interferometer,'' also known as a "non-linear Mach-Zehnder\cite{yurke19862,chekhova2016nonlinear,burlakov1997interference}. This experiment can be thought of as a nonlinear adaptation of the MZI, which has two beam splitters, whereas the SU(1,1) has two nonlinear media, $Q_1$ and $Q_2$. The object with transmission function $T=|T|e^{i\gamma}$ is placed in the idler mode between the crystals. At the output modes the two photon state can be written as
 \begin{equation}
     |\psi\rangle=\frac{(|T|e^{\gamma+\phi}+1)|1\rangle_S|1\rangle_I+\sqrt{1-|T|^2}|1\rangle_S|1\rangle_0}{2}.
 \end{equation} 
 The (singles) counting rate at detectors placed on either output path, is therefore
 \begin{equation}\label{eq:herzog-counts}
\mathcal{R}_{S/I}=\frac{1+|T|\cos(\phi+\gamma)}{2},    
 \end{equation}
 giving the intererence visibility $\mathcal{V}_{Y}=|T|$, just as in the case of the ZWMI (Eq.\ref{vis-zwm}). 
As the interferometer path of the signal, idler or pump is adjusted, both signal and idler count rates oscillate, a clear manifestation of interference. The very unique feature of this particular interferometer is that the single photon output ports are in phase with each other, though out of phase with the laser output port. That means that if maximum(minimum) counts are observed in output mode $S$, maximum(minimum) counts are simultaneously observed in output mode $I$. This leads to the curious phenomenon of \textit{Frustrated Down-Conversion}, analysed in Ref.\cite{Herzog}. If one introduces which path information in the signal or idler paths between the crystal, for example by unaligning the modes, interference is reduced or even lost in both signal and idler modes, showing complementarity between {\it welcher weg} ({\it which-path}) information and interference visibility. 
\par
The experiments described above and illustrated in Figs. \ref{fig:basicmandel} and \ref{fig:herzog} can
be viewed as ``quantum eraser'' experiments \cite{zajonc1991quantum, kwiat1994three}.
In the ZWMI interferometer (Fig.\ref{fig:basicmandel}) after the crystals but before the final beam splitter no interference would appear in either of the detection modes since the path itself marks which crystal experienced the the photon-generating down-conversion, however \emph{after} the final beam splitter this information is erased (a photon in either mode could have come from either crystal) and thus interference appears. In the SU(1,1) interferometer (Fig.\ref{fig:herzog}) after the first crystal but before the second crystal no interference would appear in either of the detection modes since the path itself marks which crystal experienced the the photon-generating down-conversion, however \emph{after} the second crystal this information is erased and interference is observed. 
Also note that, unlike in interferometers such as Mach–Zehnder, Michelson and Sagnac, where only a single phase shift is possible, in the nonlinear interferometers we discussed here, one can independently change the phases of the pump, of the signal field, and of the idler fields, and usually these have different frequencies.
\par
\begin{figure}[t]
\centering
\includegraphics[width=0.8\linewidth]{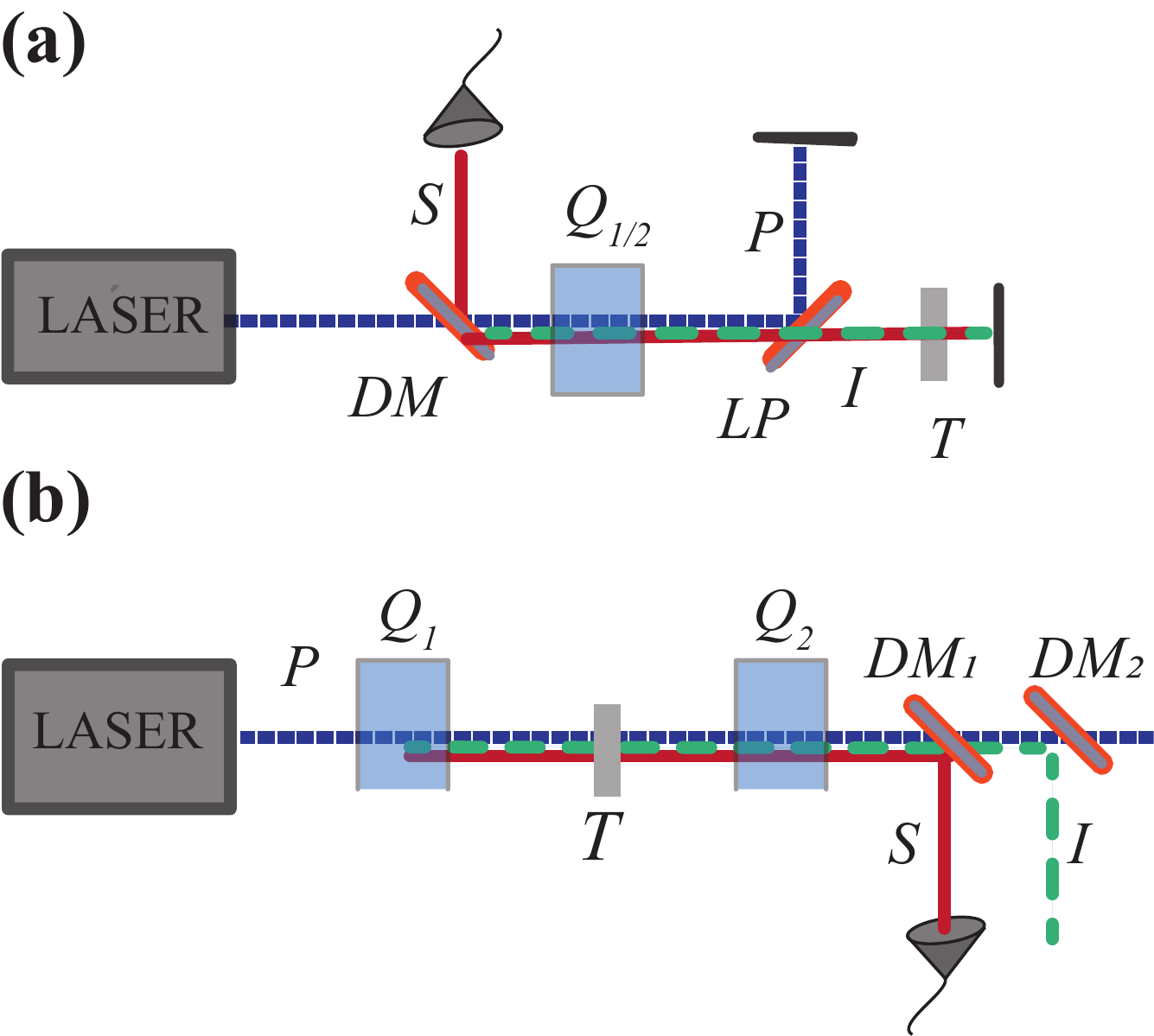}
\caption{These are two alternative versions of the SU(1,1) interferomter. In (a) the same crystal acts as sources $Q_1$ and $Q_2$, as pump, signal and idler are reflected back into that crystal. the Dichroic Mirror (DM) only reflects the signal field and the Long Pass (LP) only reflects the pump. In (b) we show a setup for collinear non-degenerate downconversion, where both signal and idler pass through the sample. Dichroic mirror $DM_1$ separates the signal $S$ from the other fields, and dichroic mirror $DM_2$ seperates the idler $I$ from the pump.}
\label{fig:herzogalternative}
\end{figure}

 Notice that in the setup in Fig.\ref{fig:herzogalternative}a signal and idler go though the object, and in Fig.\ref{fig:herzogalternative}a  all three fields, signal, idler and pump, go through the imaged object. In that case the equations in the theory sections must be adapted accordingly. In addition, we have shown in this tutorial the imaging due to light transmitted through an object, but it is trivial to adapt the theory to the case of a reflective object, which is the case of the Optical Coherence Tomography setup (Fig.\ref{fig:OCT}b), described in subsec.\ref{subsec:OCT}.
 \par 
\par
 It turns out that, in addition to the ZWMI and the SU(1,1) configurations, there is a whole range of "nonlinear interferometers", \textit{i.e}  interferometric architectures that involve two or more nonlinear sources\cite{yurke19862, chekhova2016nonlinear}. nonlinear interferometers have proved to be interesting and useful for imaging, spectroscopy, metrology and other applications\cite{Hudelist:2014di, Kalashnikov:2016cl, Iskhakov:2015gz, sven-microscopy, sven-oct, paterova2020hyperspectral, Torres-frequencycorrelation, Torres2018}. In this tutorial we will explain how to build and model some of these. 
 \section{Brief introduction to the main quantum optics states and operators}\label{sec:operators}

The basic description of interferometric devices is given in terms of the so-called creation and annihilation operators of the quantum-optical field, which are the ladder operators that work on the photon number degree of freedom as
\begin{eqnarray}
|n\rangle=\frac{(\hat{a}^{\dagger})^{n}}{\sqrt{n!}}|0\rangle,
\end{eqnarray}
where $|0\rangle$ has the standard definition of the vacuum state. A mode in a vacuum state represents a physical electromagnetic mode with only the zero-point energy.
\par
A monochromatic light field in a pure state (light is most commonly found in the form of pure states, as mixed states are difficult to produce with objects that do not typically interact with each other) may exist in a quantum mechanical superposition of containing different numbers of photons. Thus when a measurement is performed on the state it is found to have a specific number of photons. Mathematically this may be expressed, for some state $|\psi\rangle$, as
\begin{eqnarray}
|\psi\rangle=\sum_{n=0}^{\infty}p_{n}|n\rangle,
\end{eqnarray} 
where the $p_{n}$'s are the probability amplitudes of the various definite photon number states. Since any quantum state may be decomposed in the number basis we can call it \textit{complete}. Also, since orthonormality includes linear independence we can state that the the basis of definite number states (also called Fock states after Vladimir Fock, the Soviet Physicist) constitute a true and complete mathematical basis.
\par
One can write the operator corresponding to the electric field in the following simplified form \cite{gerry2005introductory}
\begin{eqnarray}
\hat{E}(t,z)=E_{0}\sin(kz)[\hat{a}e^{-i\omega t}+\hat{a}^{\dagger}e^{i\omega t}].\label{quantumE}
\end{eqnarray}
We can rewrite this in terms of two dimensionless quantities called quadrature operators defined as
\begin{eqnarray}
\hat{q}&=&\frac{1}{2}(\hat{a}+\hat{a}^{\dagger}),\label{quaddef1}\\
\hat{p}&=&\frac{1}{2i}(\hat{a}-\hat{a}^{\dagger}).\label{quaddef2}
\end{eqnarray}
Inverting these equations and substituting into Eq. (\ref{quantumE}) yields
\begin{eqnarray}
\hat{E}(t,z)=2E_{0}\sin(kz)\left[\hat{q}\cos(\omega t)+\hat{p}\sin(\omega t)\right].
\end{eqnarray}
From this it is clear that the two quadrature operators are always $\pi/2$ out of phase, and thus always in different quadratures (hence the name).
\par
It is standard to define an uncertainty relation using the generalized Heisenberg uncertainty principle for non-commuting operators $\Delta A\Delta B\geq\frac{1}{2}|\langle[\hat{A},\hat{B}]\rangle|$. This yields 
\begin{eqnarray}
\Delta\hat{q}\Delta\hat{p}\geq\frac{1}{4}.\label{quaduncert}
\end{eqnarray}
If the equality in this expression is obtained we have a so-called ``minimum uncertainty state''.

\begin{figure}
\centering
    \includegraphics[width=0.7\linewidth]{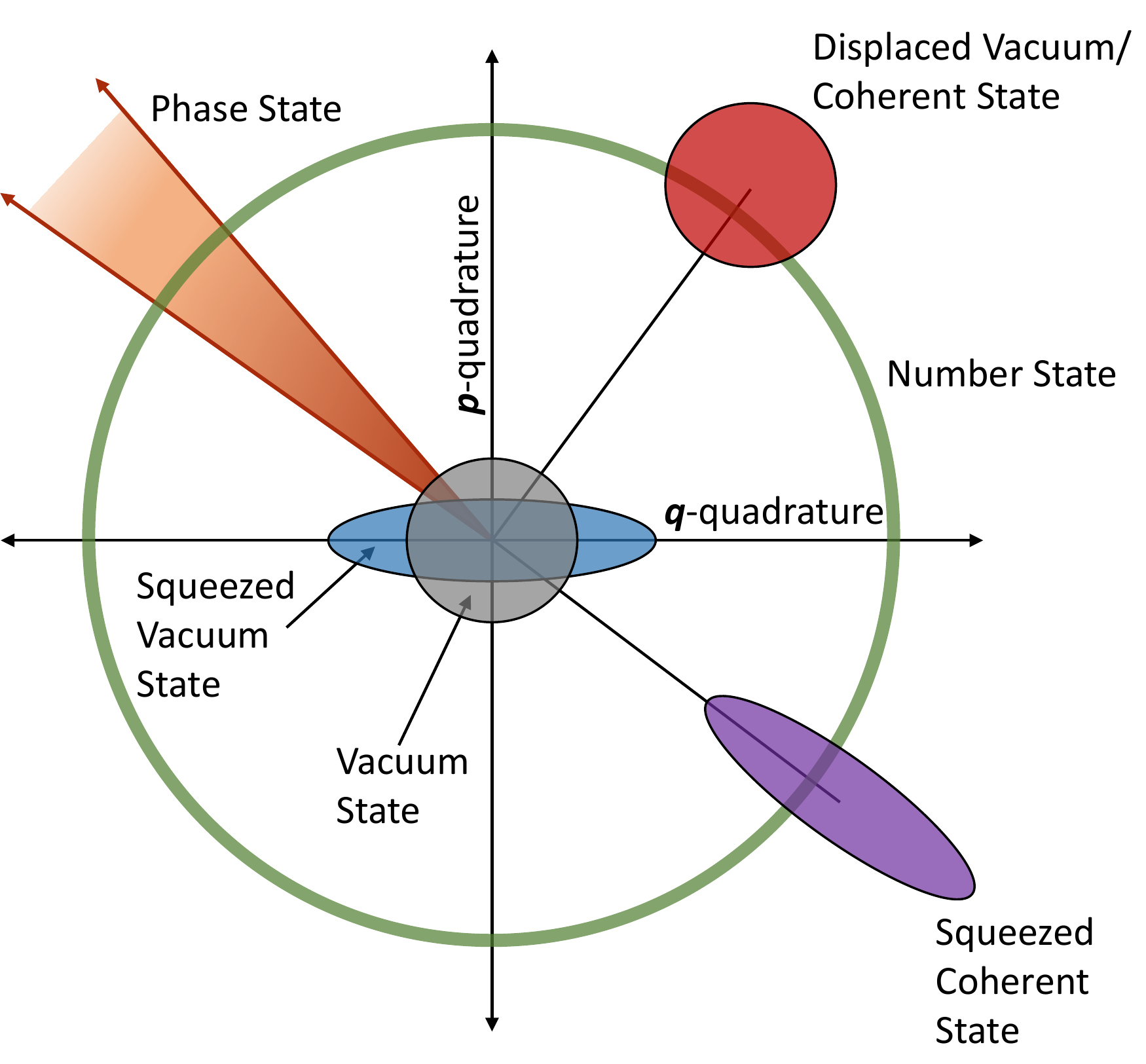}
    \caption{\textbf{Quadrature Diagram:} A configuration space defined by the two quantum-optical quadratures $\hat{p}$ and $\hat{q}$. Five states are shown: the fundamental vacuum state (gray), a displaced vacuum state / coherent state (red) \--- both symmetric minimum uncertainty states, a number state (green) with completely defined intensity and completely undefined phase, a phase state with opposite uncertainties from the number state (orange-brown, note \--- in principle this state should infinitesimally thin, but then you couldn't see it); and two squeezed states, a vacuum state squeezed in the quadrature direction (blue), and a coherent state squeezed in the phase direction (purple).} 
    \label{fig:quadrature}
\end{figure}

It's helpful to represent quantum-optical states as shapes in the quadrature space defined by Eqs.(\ref{quaddef1},\ref{quaddef2}). Specific single-mode states are represented in this diagram by their variances (the physical extent and shape of the states) and by their average values (coordinate position of the states). The diagram can also be interpreted as an intensity-phase plot using the circular-polar coordinate system (with radial distance as intensity, and angle as phase). States rotate about the origin as they evolve with time. So for example the coherent state will trace out sinusoids in its quadrature values. 
A number of such visualisations can be seen in Fig.\ref{fig:quadrature}. The shapes represent the quadrature variances as defined in Eq.(\ref{quaduncert}), technically they can be thought of as a slice through the Wigner distribution at half-maximum. The states themselves with be briefly described, and referenced, in the rest of this section and the next.   

Firstly, there is the vacuum state, representing an optical mode unoccupied by photons. Since in the quantum regime the zero point energy is always present, this state still has finite quadrature variances. Secondly, and vital for the description of interferometers, are the \emph{coherent} and \emph{squeezed} states. Coherent states can be defined in three different ways: as the state that obtains the equality in the uncertainty relation with $\Delta p=\Delta q$, as a displaced vacuum state with the displacement operator defined as
\begin{eqnarray}
\hat{D}(\alpha)\equiv e^{(\alpha\hat{a}^{\dagger}-\alpha^{*}\hat{a})}
\end{eqnarray}
which displaces the state it operates on in quadrature space by an amount $\alpha$ and generates a coherent state as $\hat{D}(\alpha)|0\rangle=|\alpha\rangle$ (see Fig.\ref{fig:quadrature}), and as eigenstates of the annihilation operator $\hat{a}|\alpha\rangle=\alpha|\alpha\rangle$. For optical fields the definitions are equivalent. Coherent states are also the most ``classical'' in the sense that their electric fields have a coherent waveform resembling a classical harmonic oscillator and that they are unaffected by the removal of a quanta of light (i.e. that they are eignstates of annihilation).

The squeezed states take their name from the uncertainty relation shown above. If $p$ and $q$ are associated with coordinates in quadrature space then $\Delta p$ and $\Delta q$ can be thought of as distances and the product $\Delta p \Delta q$ as an area. If the equality is obtained in the uncertainty relation this sets a specific minimum area of uncertainty in quadrature space. However though this area may be not reduced beyond this minimum its shape may be altered, allowing a reduction of the uncertainty along one quadrature at the cost of increasing it along another. This ``squeezing'' of the area of uncertainty gives squeezed states their name. Quantum-optical states may also be squeezed along other bases (for example photon number and phase). In Fig.\ref{fig:quadrature} two squeezed states are displayed, one is a vacuum state that has been quadrature-squeezed, and the other is a coherent state that has been phase-squeezed. 

Mathematically the squeezing operation on a single quantum-optical mode may be described as
\begin{align}\label{sqop}
\hat{S}(\xi)=\mathrm{exp}\left[\frac{1}{2}\left(\xi^{*}\hat{a}^{2}-\xi\hat{a}^{\dagger 2}\right)\right],
\end{align}
where $\xi$ is known as the squeezing parameter and quantifies the amount of squeezing. 

Acting on a mode operator this produces
\begin{eqnarray}
\hat{S}^{\dagger}(\xi)\hat{a}\hat{S}(\xi)&=&\hat{a}\cosh(r)-\hat{a}^{\dagger}e^{i\theta}\sinh(r),\\
\hat{S}^{\dagger}(\xi)\hat{a}^{\dagger}\hat{S}(\xi)&=&\hat{a}^{\dagger}\cosh(r)-\hat{a}e^{-i\theta}\sinh(r),
\end{eqnarray}
where we have used the polar decomposition $\xi=re^{i\theta}$. Equivalently we can have the squeezing operator acting on a state. Here the vacuum,
\begin{eqnarray}
\hat{S}(\xi)|0\rangle\equiv|\xi\rangle=\frac{1}{\sqrt{\cosh(r)}}\sum_{n=0}^{\infty}(-1)^{n}\frac{\sqrt{(2n)!}}{2^{n}n!}e^{in\theta}[\tanh(r)]^{n}|2n\rangle.
\end{eqnarray}
 So the squeezing operation adds a superposition of even numbers of photons to the vacuum. 
\begin{comment}
It is possible to define a two-mode squeezed vacuum state where the photons exist in separate modes. The action of two-mode squeezing is represented by
\begin{eqnarray}
\hat{S}_{2}(\xi)=\mathrm{exp}\left(\xi^{*}\hat{a}\hat{b}-\xi\hat{a}^{\dagger}\hat{b}^{\dagger}\right),
\end{eqnarray} 
which transforms mode operators and the vacuum state by 
\begin{eqnarray}
\hat{S}_{2}^{\dagger}(\xi)\hat{a}\hat{S}_{2}(\xi)&=&\hat{a}\cosh(r)-e^{i\theta}\hat{b}^{\dagger}\sinh(r),\\
\hat{S}_{2}^{\dagger}(\xi)\hat{b}\hat{S}_{2}(\xi)&=&\hat{b}\cosh(r)-e^{i\theta}\hat{a}^{\dagger}\sinh(r),\\
\hat{S}_{2}(\xi)|0,0\rangle\equiv|\xi_{2}\rangle&=&\frac{1}{\cosh(r)}\sum_{n=0}^{\infty}(-1)^{n}e^{in\theta}[\tanh(r)]^{n}|n,n\rangle.
\end{eqnarray}
Here $\hat{a}$ and $\hat{b}$ represent different (usually physical) modes, as do the two positions in the state vector.
\end{comment}
\par
Since squeezed light is a fundamental resource in nonlinear interferometry it's important to see how it is produced physically. Inside of certain materials, where light fields spatially coincide, their component electric (and magnetic) fields may no longer simply add as vectors like they do in free space. Consider the expansion of the polarization of a general dielectric material 
\begin{eqnarray}
\frac{P_{i}}{\epsilon_{o}}=\sum_{j}\chi_{ij}^{(1)}E_{j}+\sum_{jk}\chi_{ijk}^{(2)}E_{j}E_{k}+\sum_{jkl}\chi_{ijkl}^{(3)}E_{j}E_{k}E_{l}+...
\end{eqnarray}
The polarization $P$ represents how a dielectric material reacts to the presence of electric fields. The index $i$ runs over the three-dimensional vector components. The constant $\chi_{ij}^{(1)}$ is called the first order (or linear) susceptibility and it is a complex vector constant. Likewise $\chi_{ijk}^{(2)}$ is the second order susceptibility (a tensor constant), and on and on in that manner. Most materials have only a non-negligible $\chi^{(1)}$. In this case as an electric field interacts with the material it induces an oscillating dipole moment, which in turn creates an oscillating electric field, and so on. Therefore the light propagates through the material with a dispersion and absorption determined by the real and imaginary parts of $\chi^{(1)}$, respectively. 

However, now consider the case where the other terms in this series are non-negligible. Take the case of a material with a large $\chi^{(2)}$, for example. Take an electric field that has two separate frequency components 
\begin{eqnarray}
E(t)=E_{1}\left(e^{i\omega_{1}t}+e^{-i\omega_{1}t}\right)+E_{2}\left(e^{i\omega_{2}t}+e^{-i\omega_{2}t}\right),
\end{eqnarray}
and input it into the second term of the polarization (ignoring the tensor nature of $\chi^{(2)}$, i.e. assuming co-linearity).
\begin{eqnarray}
P^{(2)}(t)&=&\epsilon_{o}\chi^{(2)}\left[E_{1}^{2}\left(e^{2i\omega_{1}t}+e^{-2i\omega_{1}t}\right)\right.\label{dfg1}\\
& &+E_{2}^{2}\left(e^{2i\omega_{2}t}+e^{-2i\omega_{2}t}\right)\nonumber\\
& &+2E_{1}E_{2}\left(e^{i(\omega_{1}+\omega_{2})t}+e^{-i(\omega_{1}+\omega_{2}t)}\right)\nonumber\\
& &\left.+2E_{1}E_{2}\left(e^{i(\omega_{1}-\omega_{2})t}+e^{-i(\omega_{1}-\omega_{2}t)}\right)+2E_{1}^{2}+2E_{2}^{2}\right].\nonumber
\end{eqnarray}
Here we have assumed that the electric amplitudes are real. Eq. (\ref{dfg1}) gives rise to several interesting phenomena, but we will be interested in the effect caused by the terms that oscillate as $\omega_{1}-\omega_{2}$. This tells us that, if two light beams of different frequencies, $\omega_{1}$ and $\omega_{2}$, pump a material with a large second order susceptibility (a tensor constant, $\chi^{(2)}$) a new light beam at a frequency of $\omega_{3}=\omega_{1}-\omega_{2}$ is generated. This is called ``difference frequency generation''.
Quantum mechanically the $\omega_{2}$ mode need not be populated by photons (that is, it may be in the vacuum state) for the process to occur. In this case the interpretation is that a photon from a strong beam (called the pump) splits into two daughter photons inside the optical nonlinearity. If the two daughter photons are in the same spatial and spectral modes then we can write the interaction Hamiltonian for this process as 

\begin{eqnarray}
\hat{H}_{I}=i\hbar\chi^{(2)}\left(\hat{a}^{2}\hat{b}^{\dagger}-\hat{a}^{\dagger 2}\hat{b}\right),
\end{eqnarray}

\noindent where $\hat{b}$ represents the pump mode and $\hat{a}$ represents the mode of the daughter photons. The second term expresses one photon being transformed into two, the first term is present because the Hamiltonian must be Hermitian. Suppose the pump beam is in a coherent state (this is called the parametric approximation, meaning that the pump is \textit{undepleted}: no photons are lost), then we can write for the daughter fields alone

\begin{eqnarray}
\langle\beta|\hat{H}_{I}|\beta\rangle&=&i\hbar\chi^{(2)}|\beta|\left(\hat{a}^{2}e^{i\omega_{1}t}-\hat{a}^{\dagger 2}e^{-i\omega_{1}t}\right),\\
&=&i\hbar\chi^{(2)}|\beta|\left(\hat{a}^{2}e^{i(\omega_{1}-2\omega_{2})t}-\hat{a}^{\dagger 2}e^{-i(\omega_{1}-2\omega_{2})t}\right).\nonumber
\end{eqnarray}

\noindent Where in the second line the time dependence of $\hat{a}$ and $\hat{a}^{\dagger}$ has been made explicit. Here we choose $\omega_{1}=2\omega_{2}$, so the Hamiltonian is in fact time-independent and we can write the time evolution operator for the system simply as

\begin{eqnarray}
\hat{U}(t)=e^{-i\hat{H}_{I}t/\hbar}=e^{\hbar\chi^{(2)}t|\beta|\left(\hat{a}^{2}-\hat{a}^{\dagger 2}\right)}.
\end{eqnarray}

\noindent Now compare this to Eq. (\ref{sqop}) and we see that we have the single mode squeezing operator where $\xi=2\hbar\chi^{(2)}t|\beta|$. Furthermore we could get two mode squeezing for the case where the daughter photons are not in the same modes. So we have a way to physically squeeze vacuum (and other) states.

This is the process that occurs in both Fig.\ref{fig:basicmandel} and Fig.\ref{fig:herzog} in the nonlinear crystals creating squeezed light between the two non-pump output modes, typically called ``signal'' and ``idler'' for historical reasons not relevant here.

It should be remarked that in realistic settings the Hamiltonian at different times will not commute since frequency values are not completely sharp and have some distribution. Also higher-order terms do indeed contribute. So the above is an approximation, though one that is almost-universally standard and in good agreement with experiments involving two-photon coincidence measurements. 

Now, if we take $\hat{U}(t)$ acting on the vacuum and expand it as a power series
\begin{eqnarray}
\hat{U}(t)|0\rangle=|0\rangle+\hbar\chi^{(2)}t|\beta||2\rangle+(\hbar\chi^{(2)}t|\beta|)^{2}|4\rangle+...
\label{eq:powerseries}
\end{eqnarray}
For a coherent pump $|\beta|$ which is strong we can get several of these terms. In this case one pump photon can split into two (second term), two pump photons can combine and then split into four (third term) and so on. However if the pump is not very strong only the first two terms will be non-negligible. In this case we get the state vector for spontaneous parametric down-conversion.

Now we move away from the description of the states created inside the nonlinear interferometers and look at the other components. The transformation that represents a phase shift is
\begin{eqnarray}
 \hat{a}_{2}=e^{i\phi}\hat{a}_{1}.\label{billphase}
\end{eqnarray}  The theoretical description also needs the operator transformations for beam splitters, which are given as

\begin{eqnarray}
\hat{a}_{2}=\frac{1}{\sqrt{2}}\left[\hat{a}_{1}+i\hat{b}_{1}\right], \quad \hat{b}_{2}=\frac{1}{\sqrt{2}}\left[\hat{b}_{1}+i\hat{a}_{1}\right], \label{billBS}
\end{eqnarray}

\noindent where the first two transformations are the fields after a 50:50 beam splitter, with two input fields. The field operators $\hat{a}$ and $\hat{b}$ represent the two modes (both input and output), and the subscripts 1 and 2 represent before and after the beam-splitter (input and output), respectively. The $i$ factor on the opposite mode is the result of the phase shift of the mode under reflection.

Given these transformations the propagation of the quantum state of the device from input to output is straightforward, if sometimes somewhat burdensome. There are two basic methods for analyzing the optical modes. We can take the second term in Eq.(\ref{eq:powerseries})  \--- that is assume spontaneous parametric downconversion. Then the mode operators acting on the vacuum are transformed according to the relations above and nonlinearities act on the overall state by adding creation operators to the relevant modes.

This approach is only an approximation since the full description of the nonlinearities is given by all terms in Eq.(\ref{eq:powerseries}) and the action of the operators representing all optical elements in the device. However, performing these operations on the full state is highly non-trivial. It is much more efficient to start with a target detection operator and propagate them \emph{backwards} through the device to the coherent or vacuum input states. Then all of the elements of the optical device become linear transformations on the set of all creation and annihilation operators.

With these basic building blocks we can understand how the fields, represented by operators or states or diagrams propagate through the device and we can use these as tools to understand the various effects present. In the next section we will examine one of the most common applications of interferometer: phase detection. And in the sections after that we will see how the basic single-mode description generalizes to the multi-mode case and allows the study of imaging. 

\section{Theory of Phase Metrology with Undetected Photons} \label{sec:metrology}

A typical task in interferometry is to determine the relative phase delay between two (or more) modes of the device. For example in an MZI (Fig.\ref{fig:Mach-Zehnder}) we can measure the delay between the two modes which yields information about the difference of path lengths between modes A and B (or, perhaps, the index of refraction of some intervening transparent material, or some other similar thing). In the simplest case this information can be abstracted out as a phase shift, and in fundamental studies of interferometers the actual mechanism of the shift is typically ignored. Devices are then characterized by the minimum phase shift which can be observed  \--- corresponding to the most sensitive configuration.

The fundamental limit for \emph{classical} interferometers (typically classified as those that use coherent light only) is called the ``standard quantum limit" (SQL), which itself is the combination of two effects.

The first is the radiation-pressure noise, which is a result of the light beam imparting a fluctuating amount of momentum into the mirror. Obviously, the imparted momentum causes ``jitter'', fowling up the very sensitive phase measurements an interferometer might otherwise perform. This noise increases as the power of the light increases. Conversely it decreases as the mass of the mirror increases. Much of the research into reducing radiation pressure noise is concerned with mirror stabilization. 

The second is shot noise, which is a result of the photon number fluctuations from ``shot to shot''. Shot noise \emph{decreases} as the intensity increases, in contrast to radiation-pressure noise. This is the noise source that is typically the limiting factor. In principle the mass of the mirrors in an interferometer may be made very large, such that radiation pressure becomes small when compared to the shot noise. Though in practice this may be very difficult, there are a number of systems where the dominant source of noise is indeed the shot noise and it is often the case that the terms SQL and shot noise limit (SNL) are used interchangeably. 

By using simple arguments about the statistics of coherent states the limiting case for classical nterferometry may be found to be
\begin{eqnarray}
\Delta\phi_{\mathrm{min}}=\frac{1}{\sqrt{\bar{n}}},\label{cl}
\end{eqnarray}
\noindent which is the shot noise limit on the minimum detectable phase shift. This is the best that can be done classically. The factors $\Delta\phi_{\mathrm{min}}$ and $\bar{n}$ are the minimum detectable phase shift, and average photon number, respectively.
\par
Now the obvious questions are ``Can this be improved upon using quantum resources? And, if so, what new fundamental limit constrains quantum devices?'' The answers to these questions are commonly agreed to be ``yes'' and ``the Heisenberg limit,'' respectively.

The Heisenberg limit presumes to be the absolute limit on the phase sensitivity of an interferometer. Unlike the shot noise limit in Eq.~(\ref{cl}), it makes no assumptions about the specific kind of light being used. Instead the Heisenberg limit draws upon the fundamental laws of quantum mechanics to place a bound on how accurately we may measure a light field's (or matter wave's) phase. However, it must be noted that it is not straightforward to provide a rigorous derivation of the Heisenberg limit for the phase measurement, which is accepted by all. The root of the problem lies in the definition of the phase operator. Below we provide a brief and lucid description of the issue. 
\par
A phase operator was originally introduced by Dirac in his celebrated paper on quantum theory of radiation \cite{dirac1927quantum}. To understand Dirac's approach, let us take the annihilation operator acting on a coherent state
\begin{eqnarray}
\hat{a}|\alpha\rangle=\alpha|\alpha\rangle=e^{i\phi}|\alpha||\alpha\rangle=e^{i\phi}\sqrt{\bar{n}}|\alpha\rangle.
\end{eqnarray}
The annihilation operator is a purely quantum mechanical object with no classical analogue. However, it can be decomposed into quantities that \textit{are} familiar in classical optics: the average intensity, $\bar n$, and the phase, $\phi$, of an optical field. Dirac took this to mean that the creation and annihilation operators could be factored into Hermitian observables as
\begin{subequations}
\begin{align}
&\hat{a}=e^{i\hat{\phi}}\sqrt{\hat{n}} \\
&\hat{a}^{\dagger}=\sqrt{\hat{n}}e^{-i\hat{\phi}},
\end{align}
\end{subequations}
where the second equation is obtained by simply taking the conjugate transpose of the first. He thus defined the phase operator $\hat{\phi}$. This combined with the commutation relation $[\hat{a},\hat{a}^{\dagger}]=1$ yields $[\hat{n},\hat{\phi}]=i$. 
\par
Dirac thus concluded that photon number and phase are conjugate (canonical) observables. Therefore, in order to become more certain about one, we must become less certain about the other. This relationship can be made quantitative by employing the generalized Heisenberg uncertainty principle for non-commuting operators: $\Delta A\Delta B\geq\frac{1}{2}|\langle[\hat{A},\hat{B}]\rangle|$. Using this we have $\Delta n\Delta\phi\geq 1/2$. So in order to know as much as we can about the phase, we must reduce by as much as possible our knowledge of the number of photons in the field. Since we are concerned with fundamental limits we will take the case where photon number is as uncertain as possible: when $\Delta n=\bar{n}/2$. The uncertainty can not be made any larger than this because then there would be a non-zero probability of detecting a negative number of photons in the field (which is physically meaningless). Therefore, we get the following expression for the Heisenberg limit:  
\begin{eqnarray}
\Delta\phi_{\mathrm{min}}=\frac{1}{\bar{n}}.\label{heilim}
\end{eqnarray}
However, there is a flaw in this argument. The problem exists in the definition of the phase operator that was later shown to be non-Hermitian \cite{louisell1963amplitude,susskind1964quantum}. In fact, if we attempt to write down an eigenstate for this phase operator we can visualize the issue.  
\par
Eigenstates of the number operator (representing a light field with an exactly-known number of photons but a completely undefined phase) are well-defined as a ring in quadrature space with a diameter equal to the intensity of the field and a finite area (and thus energy), see Fig.~\ref{fig:quadrature}. However, in quadrature space, a phase eigenstate is a wedge radiating out from the origin to infinity (all of the space that exists at a particular angle). Such a state has infinite area and thus infinite energy, and thus is not normalizable (see Fig.~\ref{fig:quadrature} again).  
\par
Since the problem with Dirac's phase operator was pointed out, there have been numerous discussions and proposals on this issue (see, for example, \cite{susskind1964quantum,barnett1989hermitian,Noh-1991-Phase,ou1997fundamental}). Furthermore, there exists another approach to understand the achievable precision of phase measurement from the perspective of quantum estimation theory \cite{pirandola2018advances,braun2018quantum}. Nevertheless, the Heisenberg limit ($\Delta\phi_{\mathrm{min}}=1/\bar{n}$) is widely used and commonly regarded as an approximate bound, in the limit of high photon number \cite{louisell1963amplitude,ou1997fundamental}. The Heisenberg limit remains a useful and common goalpost for studies in interferometry.

Now suppose we wish to consider a \emph{specific} device that probes the abstract phase shift by imprinting it on some measurable quantity. Mathematically this means we have some quantity that is a function of the phase $M(\phi)$. So we ask the question ``Given that we are measuring $M$ what is the smallest change we can detect in $\phi$?'' To answer this question start with the Taylor series expansion of a function $M$ with the variable as $\phi$ about a point $\phi_{o}$

\begin{eqnarray}
M(\phi)=M(\phi_{0})+(\phi-\phi_{0})\left.\frac{\partial M}{\partial \phi}\right|_{\phi\rightarrow \phi_{0}}+\quad ...
\end{eqnarray}

\noindent Where $\partial(\phi-\phi_{0})=\partial\phi$ since $\phi_{0}$ is a constant. The smallest detectable phase shift would be equal to the smallest we could make $\phi-\phi_{0}$. Since we are considering this quantity to be very small, we can truncate the series after the second term and recast the above as

\begin{eqnarray}
\frac{M(\phi)-M(\phi_{0})}{\left.\frac{\partial M}{\partial \phi}\right|_{\phi\rightarrow \phi_{0}}}=\phi-\phi_{0}=\Delta\phi_{\mathrm{min}}.\label{avg1}
\end{eqnarray}

\noindent If we take $\phi_{0}$ to be the \textit{average} value of the phase then $M(\phi)-M(\phi_{0})$ gains the interpretation of being the variance of $M$ for a \emph{single} measurement, as $M(\langle\phi\rangle)=\langle M(\phi)\rangle$ \--- and the derivative becomes the derivative with respect to phase of $\langle M\rangle$. However, we want the statistically averaged variance for a \emph{series} of measurements (the standard deviation) of $M$, so we take $(M-\langle M\rangle)^{2}$ and average it (which is the standard deviation squared), which yields $\langle M^{2}\rangle-\langle M\rangle^{2}$.

%\begin{eqnarray}
%\left.\frac{\partial M}{\partial \phi}\right|_{\phi\rightarrow \phi_{0}}\rightarrow\frac{\partial \langle M\rangle}{\partial \phi},
%\end{eqnarray}

%\begin{eqnarray}
%(\Delta\phi_{\mathrm{min}})^{2}=\frac{(M-\langle M\rangle)^{2}}{\left(\frac{\partial \langle M\rangle}{\partial \phi}\right)^{2}}.\label{phimin2}
%\end{eqnarray}

%\begin{eqnarray}
%\langle(M-\langle M\rangle)^{2}\rangle&=&\sum (M-\langle M\rangle)^{2}P(M),\nonumber\\
%&=&\sum M^{2}P(M)-2\langle M\rangle\sum MP(M)+\langle M\rangle^{2}\sum P(M),\nonumber\\
%&=&\langle M^{2}\rangle-\langle M\rangle^{2},\nonumber
%\end{eqnarray}

%\noindent where $P(M)$ is the probabilistic weight function. Now substituting this back into Eq. (\ref{phimin2}) and taking the square root of the expression

Then, squaring both sides of Eq.(\ref{avg1}), making the substitutions above, and identifying $\langle M^{2}\rangle-\langle M\rangle^{2}$ as the square of the variance we arrive at

%\begin{eqnarray}
%\Delta\phi_{\mathrm{min}}=\frac{\sqrt{\langle M^{2}\rangle-\langle M\rangle^{2}}}{\left|\frac{\partial \langle M\rangle}{\partial \phi}\right|}.\nonumber
%\end{eqnarray}

\begin{eqnarray}
\Delta\phi_{\mathrm{min}}=\frac{\Delta\hat{M}}{\left|\frac{\partial \langle \hat{M}\rangle}{\partial \phi}\right|}.\label{minimumfinal}
\end{eqnarray}

\noindent Where we have promoted $M$ to a quantum mechanical observable, and taken the root of both sides. This is the minimum detectable phase shift. To calculate this we need both a choice of measurement operator, and the quantum-mechanical state that operator works on (in order to take the expectation values).

Now, recall the limit of a standard Mach-Zehnder Interferometer (MZI) with coherent light input \--- Eq.(\ref{cl}). We wish to use this new formula to calculate the sensitivity of this device to changes in the abstract phase $\phi$. We need the relationship between the input modes and the output modes given by Eq.(\ref{billBS}). These will allow us to write the operators at the detection end of the MZI in terms of the operators at the input end. We then take the expectation values of these operators at the input end. 

The most important question, with regards to our sensitivity formula and the MZI, is the choice of the detection scheme $\hat{M}$. Which for a MZI is the difference of the intensities at the bright-port and the dark-port 

\begin{eqnarray}
\hat{M}=\hat{b}_{f}^{\dagger}\hat{b}_{f}-\hat{a}_{f}^{\dagger}\hat{a}_{f}.
\end{eqnarray}

\noindent Where the subscript indicates that these operators act on the final state. This corresponds to an intensity difference measurement between modes A and B in Fig\ref{fig:Mach-Zehnder}. Using this information and Eq. (\ref{minimumfinal}) we find for this setup $\Delta\phi_{\mathrm{min}}=1/|\alpha|=1/\sqrt{\bar{n}}$. Which is, unsurprisingly, the shot noise limit in Eq.(\ref{cl}).

An analysis of the phase sensitivity has also been recently performed for the ZWMI by some of us and others \cite{miller2021versatile}. Several interesting effects are found. Firstly, as might be expected for a ``highly quantum'' device, the minimum detectable phase shift reaches below the classical bound of the shot noise limit, meaning that the device is ``super-sensitive''. This effect is maintained regardless of gain regime (at least in principle). Furthermore, when the initial crystal $Q_1$ is seeded with a strong coherent (laser) beam the sensitivity is further increased (``boosted'') into the bright-light regime while still maintaining some aspects of the super-sensitive scaling. Though the general equations produced by this calculation are very large, a simple case of the minimum detectable phase-shift (squared) can be presented for coherent light injection into one of the modes (in this case the one that does not pass through the sample, corresponding to mode $S_{1}$ in Fig.\ref{fig:basicmandel}) and intensity difference subtraction measurement between the two detected modes (modes $S_{1}$ and $S_{2}$ in the same), when the gains are very large (and equal to each other), and with the probe phase (and all other phases) set to zero:  

\begin{eqnarray}
\Delta\phi_{\mathrm{min}}^2=\frac{e^{-2r}}{4(1+\beta^{2})}.
\end{eqnarray}

\noindent Note that this equation is not optimal \--- rather it is presented because of its tractability. (for detailed discussion see Ref.\cite{miller2021versatile}). From this equation it is clear that both the squeezing due to the nonlinearities and the coherent light injection improve the sensitivity. The improvement is exponential for the gain and inverse-squared for the coherent light injection. 

The SU(1,1) interferometer, Fig.\ref{fig:herzog}, has become a commonly-studied device \cite{yurke19862,Herzog,plick2010coherent,hudelist2014quantum,oureview} due to the fact that it allows super-sensitive detection with bright light. Recently these interferometers have also been modified so that a MZI is nested inside \cite{du20202}, which has some conceptual similarity to the ZWMI configuration in the sense that both beam-splitter and squeezing operations are performed. 

\begin{figure}[t]
\centering
\includegraphics[width=\linewidth]{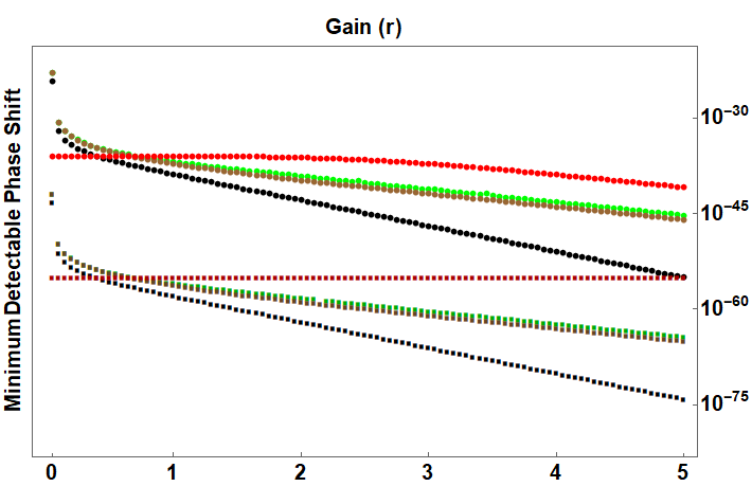}
\caption{\textbf{Metrology with undetected photons.} Figure taken from \cite{miller2021versatile}. The minimum detectable phase-shift squared of several fair comparison interferometric setups and detection schemes as a function of gain (of the first crystal for ZWMI and of \emph{both} crystals for SU(1,1)). Here we display the boosted ZWMI setup with intensity detection at mode $B$ (green), intensity difference detection between modes $B$ and $C$ (brown), the boosted SU(1,1) setup (black), and a standard coherent-light-seeded MZI with the extra light needed to create the aforementioned squeezings added to the initial input (red). The latter is equivalent to the shot-noise limit. All other parameters are numerically optimized at each point. The circular points (upper set) represent injected coherent light of about the same intensity as would be needed for a high-gain non-linearity, and the square points (lower set) represent a much-brighter coherent input.}
\label{fig:logplot}
\end{figure}

An argument against the use of all non-linear interferometers could be paraphrased as ``If we need a bright laser to pump the non-linearity, wouldn't it just be better to use that bright light in an MZI?''. To confront this criticism, any light that is used to pump a non-linear source is added to the MZI ``light budget'', making the comparison ``fair''. In figure  Fig.\ref{fig:logplot} the phase sensitivity of the ZWM, SU(1,1) and MZ interferometers are compared. This graph uses the concept of a ``fair comparison state''. We see that, though the SU(1,1) configuration performs best, the ZWMI type also displays super-sensitivity, beating the MZI after even a modest non-linear gain.

Surprisingly, this super-sensitive phase detection is available regardless of which of the two initial input modes is ``boosted'' into the bright-light regime (corresponding to injecting  coherent light either into mode $S_{1}$ or into mode $I$ in Fig.\ref{fig:basicmandel}). Therefore one can choose not to shine the extra laser through the sample and achieve the same sensitivity increase as if it had been. Likewise, one can shine the light through the sample, into the mode that is discarded and thus avoiding hitting the detectors. This technique could prove very useful in cases where either the detectors or sample is sensitive to bright coherent light. Furthermore, when contrasted with the SU(1,1) configuration, which requires adaptive intensity measurements (where the signal is produced by summing the intensities of the two output modes) or homodyning for detection, the ZWMI geometry uses intensity subtraction, making it a more stable and thus at least in some cases more experimentally desirable. 

The ZWMI has also been studied from the perspective of the signal-to-noise ratio. In Ref.\cite{giese2017phase} the authors theoretically study a ZWMI with a variable-field transmittance beam splitter inserted in the idler path between the two non-linear sources, with different pump intensities for the non-linear sources as well. They look at the signal-to-noise and visibility of the output as a function of the gain in the non-linearities and the field transmittance of the aforementioned beam splitter, paying special attention to the qualitative and quantitative difference between various regimes of gain. They also find that the visibility of the system may be optimized by proper choice of the field transmittance of the beam splitter.

So far we have examined only ``single mode'' descriptions of the ZWMI and the SU(1,1). That is, we do not take into account the real momentum and frequency distributions. We have assumed that fields can differ only in path. In the following section we build on the previous to create a more complete picture, especially as it relates to imaging.

\section{Theory of Imaging with Undetected Photons} \label{sec:rig-theory}

Amplitude and phase imaging using a Zou-Wang-Mandel interferometer was introduced in \cite{Lemos:2014p346} using collinear SPDC, as shown in Fig.\ref{fig:basicmandel}(b). Three objects were placed in the undetected idler arm with wavelength $1550$ nm: A cardboard cut-out, an etched fused silica plate and an etched silicon plate. The images of these objects were retrieved in the interference pattern of the combined $810$ nm signal field using an Electron Multiplying Charged Coupled Device (EMCCD), although a scientific Complementary metal-oxide-semiconductor (sCMOS) camera could have been used instead. Indeed, there is no requirement of single photon detection, although it is important to have a camera or spatial scanning detector that has low-light sensitivity at the signal wavelength. 
\par
In this section, we present a rigorous theoretical description of the image formation in the Zou-Wang-Mandel interferometer. In order to have a thorough idea of the imaging we must consider the multi-mode structure of optical fields. 
\par
Quantum Imaging with Undetected Photons (QIUP) relies on transverse spatial correlations between signal and idler photons. Although SPDC photon pairs in general exhibit transverse spatial entanglement \cite{walborn10}, we will see that \textit{spatial entanglement is in general not necessary for imaging via induced coherence}. In the first experiment, lenses in the idler path between the crystals (not shown in the figures) were used such that the object was in the far field relative to $Q_1$ and $Q_2$ \cite{Lemos:2014p346,Lahiri2015a}. The object was imaged onto the camera also using lenses in the signal fields. In that case, the imaging was enabled by momentum correlation between the twin photons. Alternatively, the object and the camera can be placed in the near field (source plane) relative to the twin photon sources and, in this case, the imaging is enabled by the position correlation between twin photons \cite{viswanathan2021position}. (Evanescent fields play no role in this configuration and, therefore, must not be confused with the conventional near-field imaging.) Images generated in these two cases have distinct features. We discuss the two configurations separately in Secs. \ref{sec:rig-theory}\ref{subsec:momcorr-img} and \ref{sec:rig-theory}\ref{subsec:poscorr-img}. We stress here that the theory is not restricted to twin photons generated by SPDC; it applies to spatially correlation twin photons generated by any source.  

\subsection{Multi-mode twin photon states}\label{subsec:gen-q-state}
 Throughout the analysis we assume that photons propagate as paraxial beams and are always incident normally on both the object and the detector. Under these assumptions, the two-photon quantum state can be written as (see, for example, \cite{walborn10})
\begin{equation} \label{state-spdc}
|\widetilde{\psi}\rangle = \int d\textbf{q}_{s}  \ d\textbf{q}_{I} \ C(\textbf{q}_{s}, \textbf{q}_{I}) |\textbf{q}_{s}\rangle_{s} |\textbf{q}_{I}\rangle_{I},
\end{equation}
where $|\textbf{q}_{s}\rangle_{s} \equiv \hat{a}_{s}^{\dagger}(\textbf{q}_{s})|vac\rangle$ denotes a signal photon Fock state labeled by the transverse component $\textbf{q}_{s}$ of the wavevector $\textbf{k}_{s}$. Similarly, $|\textbf{q}_{I}\rangle_{I}\equiv \hat{a}_{I}^{\dagger}(\textbf{q}_{I})|vac\rangle $ denotes an idler photon Fock state labeled by the transverse component $\textbf{q}_{I}$ of the wavevector $\textbf{k}_{I}$. The complex quantity $ C(\textbf{q}_{S}, \textbf{q}_{I})$ ensures that $|\psi\rangle$ is normalized, i.e.,
\begin{equation} \label{C-fn-norm}
\int d\textbf{q}_{s}  d\textbf{q}_{I}  |C(\textbf{q}_{s}, \textbf{q}_{I})|^{2} = 1.
\end{equation} 
\par
Imaging enabled by both momentum correlation (Sec. \ref{subsec:momcorr-img}) and position correlation (Sec. \ref{subsec:poscorr-img}) can be described by the quantum state given by Eq. (\ref{state-spdc}). Such a quantum state is usually generated by SPDC at a nonlinear crystal. However, the theoretical analysis applies to any source that can generate such a state.
\par
\begin{figure}[t]
\centering 
\includegraphics[width=\linewidth]{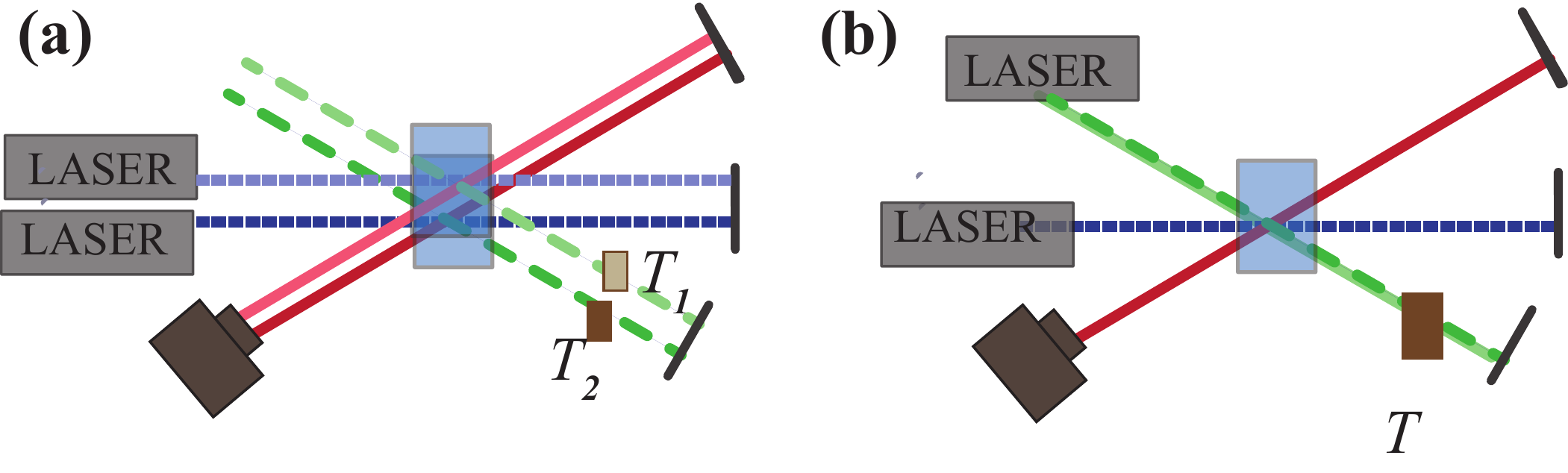}
\caption{\textbf{Classical transverse spatial correlations are sufficient for imaging}.(a) Two spatially separated interferometers can together produce an image on the camera. Transverse spatial entanglement is not necessary for QIUP. (b)In classical imaging with undetected light \cite{Cardoso2017},
two spatially separated interferometers can together produce an image on the camera. Transverse spatial entanglement is not necessary for IUP.  }
\label{fig:entanglement-explanation}
\end{figure}

The state \ref{state-spdc} can be entangled, \textit{i.e.} $C(\textbf{q}_{s}, \textbf{q}_{I})\neq C_s(\textbf{q}_{s})C_i(\textbf{q}_{I})$. However, transverse spatial entanglement is not a requirement for imaging with undetected photons. Consider, for example, two separate SU(1,1)s (or two ZWMIs), like in Fig.\ref{fig:entanglement-explanation}a. One interferometer is placed close to the other such that one camera can capture at once the signal outputs of both interferometers. In the top SU(1,1)  once places a sample with field transmittance $T_1$ and in the other one places a sample with field transmittance $T_2$. An image of $T_1$ and $T_2$, and their spatial separation, can be observed on the camera. This shows that IUP requires transverse spatial correlation between signal and idler fields \textit{at the plane where the object is placed}, but spatial entanglement is not necessary. An experimental result that proves that imaging in a nonlinear interferometer is possible with only classical spatial correlations is described in ref.\cite{Cardoso2017}. In that experiment, a seeding laser was injected through the first pass of a SU(1,1) (Fig.\ref{fig:entanglement-explanation}b), where it stimulated emission into the signal mode. This laser then went through the object, was reflected on the mirror and passed again through the crystal, stimulating into the same signal mode. A picture of the object was seen on the camera. The signal and idler fields in this case are not entangled, but present classical transverse spatial correlations \cite{monken1998transfer}.

\subsection{Imaging Enabled by Momentum Correlation}\label{subsec:momcorr-img}
\subsubsection{General theory}\label{subsec:momcorr-img-thoery}
It is evident from Eq. (\ref{state-spdc}) that the joint probability density of detecting a signal photon with transverse momentum $\hbar \textbf{q}_s$ and and idler photon with transverse momentum $\hbar \textbf{q}_I$ is given by
\begin{equation} \label{prob-mom}
P(\textbf{q}_s,\textbf{q}_I) \propto |C(\textbf{q}_{s}, \textbf{q}_{I})|^{2}.
\end{equation}
This probability density characterizes the momentum correlation between the twin photons. We now show that in the far field configuration, this momentum correlation enables image formation.

\begin{figure}[ht]
\centering 
\includegraphics[width=\linewidth]{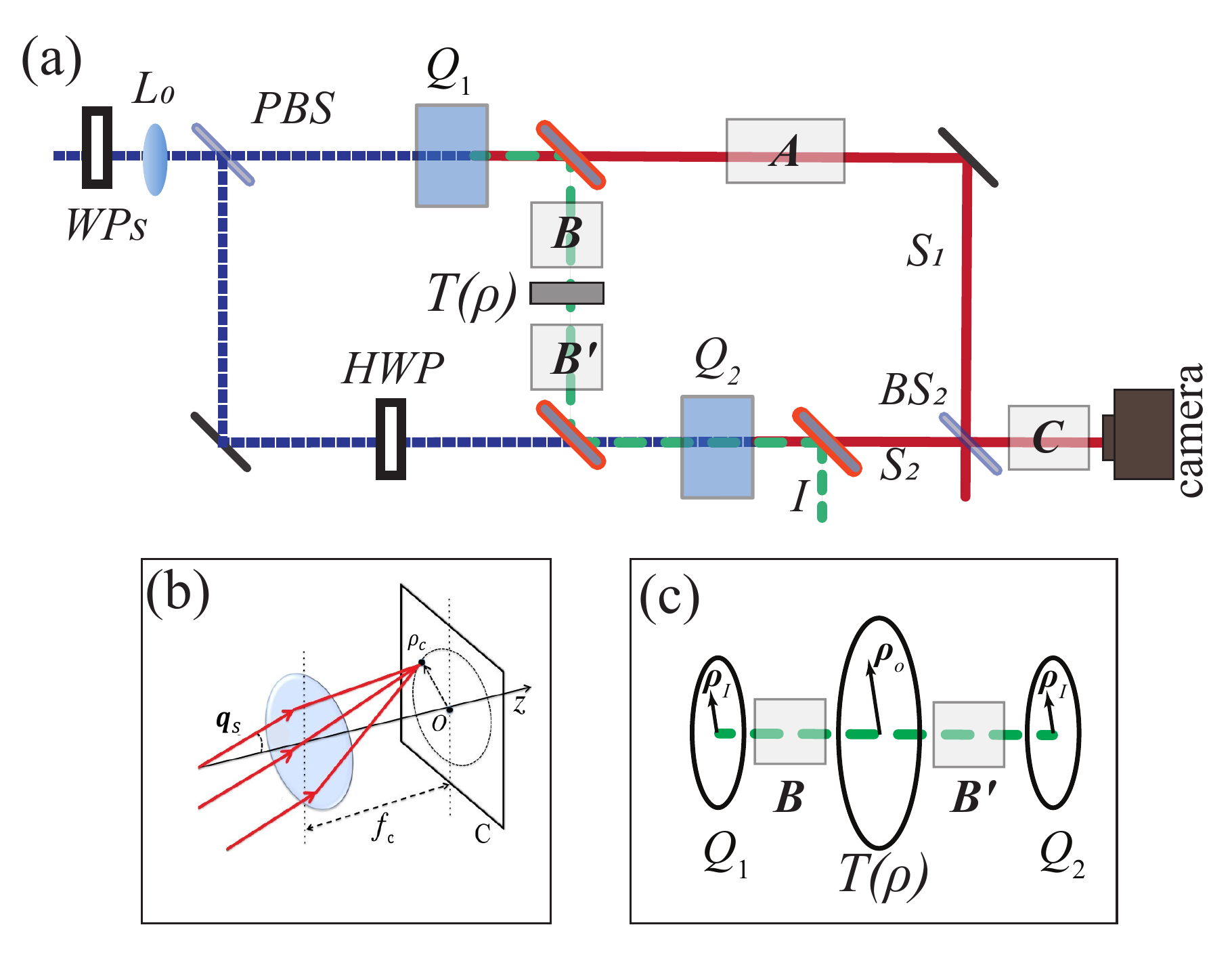}
\caption{\textbf{Quantum Imaging with Undetected Photons}. (a) In a collinear non-degenerate ZWMI\cite{Lemos:2014p412}, non-degenerate photon pairs are emitted along the propagation axis of a Laser at each source. Waveplates and a polarizing beam splitter (PBS) in the pump are used to control the relative phases and amplitudes of the two-photon states generated in sources $Q_1$ and $Q_2$. Dichroic mirrors or long pass filters can be used to separate the pump from the daughter fields after each crystal.
A lens $L_0$ is used to control the pump waist at the crystals, which affects the twin photon transverse momentum correlations, which in turn affects the image resolution, as shown in section \ref{subsec:momcorr-img-reso}. Imaging system $A$ ensures a good overlap of the combined signal fields and optical system $C$ is used to image the plane of the object with spatial features $T(\rho)$ onto the plane of the camera or scanning detector. (b) In the case of imaging enabled by momentum correlation, the optical systems $B$ and $B'$ guarantee that the object $T(\rho)$ is at the Fourier plane of sources $Q_1$ and $Q_2$. An effective positive lens with
focal length $f_c$ associates the plane on the camera with the Fourier Plane of the sources $Q_1$ and $Q_2$. A plane 
wave vector $\pmb{q}_s$ makes an angle $\theta$ with the optical axis and is focused along a circle of radius
$|\pmb{\rho}_{c}|$. (c) In the case of imaging enabled by position correlation, optical systems $B$ and $B'$ are imaging systems; a point $\boldsymbol{\rho}_{s}$ on $Q_j(j=1,2)$ is imaged onto $\boldsymbol{\rho}_{c}=M_s\boldsymbol{\rho}_{s}$ on the camera by $A$. }
\label{fig:img-schm-mc}
\end{figure} 
\par
The experimental setup is illustrated in Fig. \ref{fig:img-schm-mc}. There are two sources, $Q_1$ and $Q_2$, each of which can emit a photon pair. $Q_1$ emits the signal and idler photons into beams $S_1$ and $I_1$, respectively. Likewise, $S_2$ and $I_2$ represent the beams into which the signal and idler photons are emitted by $Q_2$. The two sources ($Q_1$ and $Q_2$) almost never emit simultaneously and almost never produce more than two photons individually. Furthermore, the two sources emit coherently. Under these circumstances, the quantum state of light generated by the two sources is given by the superposition of the states generated by them individually, i.e., by
\begin{align} \label{biphoton-state}
|\psi\rangle =& \int d\textbf{q}_{s} d\textbf{q}_{I} ~ C(\textbf{q}_{s}, \textbf{q}_{I}) \nonumber \\ & \quad \times \left[\alpha_{1}\hat{a}^{\dagger}_{s_{1}}(\textbf{q}_{s}) \hat{a}^{\dagger}_{I_{1}}(\textbf{q}_{I}) +\alpha_{2}\hat{a}^{\dagger}_{s_{2}}(\textbf{q}_{s}) \hat{a}^{\dagger}_{I_{2}}(\textbf{q}_{I}) \right] |vac \rangle ,
\end{align}
where $\alpha_{1}$ and $\alpha_{2}$ are complex numbers satisfying the condition $|\alpha_{1}|^{2} + |\alpha_{2}|^{2} = 1$, and $|vac \rangle$ represents the vacuum state.
\par
As shown in Fig. \ref{fig:img-schm-mc}, lens systems are used to place both the object and the camera in the far field (Fourier plane) of the sources. These lens systems must also ensure that source $Q_1$ is imaged onto $Q_2$ and the camera is on the image plane of the object. For the best possible alignment of idler beams, it is important that $Q_1$ is imaged onto $Q_2$ with unit magnification. Here, we present a generic treatment which does not consider any specific lens systems and provides a full understanding of the imaging mechanism. The theory can certainly be tailored to specific systems and this may only result into incremental differences such as a different sign of the image magnification. 
\par
A lens system $B$ that places the object in the Fourier plane of $Q_1$ can be effectively modeled by a single positive lens, where $Q_1$ and the object are located on the back and front focal planes of this positive lens, respectively. Similarly, the lens system $B'$ that places $Q_2$ on the Fourier plane of the object can be modeled by another positive lens. In order for $Q_1$ to be imaged onto $Q_2$ with unit magnification, the focal length these two lenses must be equal, and shall here be denoted \emph{effective focal length}, $f_I$. Since a transverse wave vector $\textbf{q}_I$ is focused on a point $\boldsymbol{\rho}_o$ on the object, within the paraxial approximation we obtain from lens rules
\begin{align}\label{q-obj-rel}
\boldsymbol{\rho}_{o}= \frac{\lambda_I f_I}{2\pi} \textbf{q}_I,
\end{align}
where $\lambda_I$ is the mean wavelength of the idler photon. 
\par
The interaction of the idler field with the object can be represented by the transformation made by a beam splitter and we, therefore, have the following expression \cite{Lahiri2015a}:
\begin{align}\label{align-mom}
\hat{a}_{I_{2}}(\textbf{q}_{I}) = e^{i\phi_I}\left[T(\boldsymbol{\rho}_o) \hat{a}_{I_{1}}(\textbf{q}_I)+ R(\boldsymbol{\rho}_o) \hat{a}_0(\textbf{q}_I)\right],
\end{align}
where $\phi_I$ is the phase due to propagation of the idler beam from $Q_1$ to $Q_2$, the operator $\hat{a}_0$ represents vacuum field at the unused port of the beam splitter (object), $T(\boldsymbol{\rho}_o)$ is the amplitude transmission coefficient of the object at a point 
$\boldsymbol{\rho}_o$ that is related to $\textbf{q}_{I}$ by Eq. (\ref{q-obj-rel}), and $|T(\boldsymbol{\rho}_o)|^2+|R(\boldsymbol{\rho}_o)|^2=1$. The quantity, $R(\boldsymbol{\rho}_o)$, can be interpreted the amplitude reflection coefficient at the same point while illuminated from the other side. 
\par
The quantum state of light generated by the system is obtained by combining Eqs. (\ref{biphoton-state}) and (\ref{align-mom}). It is given by 
\begin{align} \label{biphoton-st-mom-f}
|\Psi\rangle =& \int d\textbf{q}_{s} d\textbf{q}_{I} ~ C(\textbf{q}_{s}, \textbf{q}_{I}) [\alpha_1\ket{\textbf{q}_s}_{s_1} +e^{-i\phi_I}\alpha_2 T^{\ast}(\boldsymbol{\rho}_o)\ket{\textbf{q}_s}_{s_2}] \ket{\textbf{q}_I}_{I_1} \nonumber \\ & +\int d\textbf{q}_{s} d\textbf{q}_{I} ~ C(\textbf{q}_{s}, \textbf{q}_{I}) e^{-i\phi_I}\alpha_2 R^{\ast}(\boldsymbol{\rho}_o)\ket{\textbf{q}_s}_{s_2} \ket{\textbf{q}_I}_{0},
\end{align}
where $\hat{a}^{\dag}_0(\textbf{q}_I)\ket{vac}=\ket{\textbf{q}_I}_{0}$.
\par
Since the camera is placed in the combined signal field, at the far field relative to the sources, we can once again use the concept of an effective positive lens, as shown in Fig. \ref{fig:img-schm-mc}(b). Suppose that the focal length of this lens is denoted by $f_c$. Following an argument, which is similar to the one used to obtain Eq. (\ref{q-obj-rel}), we find that a point $\boldsymbol{\rho}_c$ on the camera is related to the transverse signal wave vector $\textbf{q}_s$ by the following formula:
\begin{align}\label{q-cam-rel}
\boldsymbol{\rho}_c= \frac{\lambda_s f_c}{2\pi} \textbf{q}_s,
\end{align}
where $\lambda_s$ is the mean wavelength of the signal photon. The quantized field at a point, $\boldsymbol{\rho}_c$, on the camera plane can now be represented by
\begin{align} \label{field-cam-mom}
\hat{E}^{(+)}_s(\boldsymbol{\rho}_c) \propto e^{i\phi_{s_1}}\hat{a}_{s_{1}}(\textbf{q}_s)+ e^{i\phi_{s_2}}\hat{a}_{s_{2}}(\textbf{q}_s),
\end{align}
where $\phi_{s_1}$ and $\phi_{s_2}$ are phases due to propagation of the signal beams from $Q_1$ and $Q_2$, respectively, to the camera. The single-photon counting rate (intensity) at a point $\boldsymbol{\rho}_{c}$ the camera can be determined by the standard formula $\mathcal{R}(\boldsymbol{\rho}_{c})\propto \bra{\Psi}\hat{E}^{(-)}_s(\boldsymbol{\rho}_c) \hat{E}^{(+)}_s(\boldsymbol{\rho}_c)\ket{\psi}$. It now follows from Eqs. (\ref{biphoton-st-mom-f}) and (\ref{field-cam-mom}) that
\begin{align} \label{inten-mom-gen}
\mathcal{R}(\boldsymbol{\rho}_{c})\propto \int d\textbf{q}_{I} ~ P(\textbf{q}_{s}, \textbf{q}_{I}) \left[1 + \left|T (\boldsymbol{\rho}_{o}) \right| \cos (\phi_{in} -\text{arg}\{T( \boldsymbol{\rho}_{o})\}) \right],
\end{align}
where $\phi_{in}=\phi_{s_2}-\phi_{s_1}-\phi_{I}+\text{arg}\{\alpha_2\}-\text{arg}\{\alpha_1\}$, arg represents argument of a complex number, and $\boldsymbol{\rho}_{o}$ \& $\textbf{q}_{I}$ and $\boldsymbol{\rho}_{c}$ \& $\textbf{q}_{s}$ are related by Eqs. (\ref{q-obj-rel}) and (\ref{q-cam-rel}) respectively, and we have assumed $|\alpha_1|=|\alpha_2|$ for simplicity.
\par

It is evident from Eq. (\ref{inten-mom-gen}) that the information about the object (both magnitude and phase of the amplitude transmission coefficient) appears in the interference pattern observed on the camera, even though the photons probing with the object are not detected by the camera. The presence of $P(\textbf{q}_{s}, \textbf{q}_{I})$ in Eq. (\ref{inten-mom-gen}) shows that the momentum correlation between the twin photons enables the image acquisition. For example, when there is no correlation between the momenta, $P(\textbf{q}_{s}, \textbf{q}_{I})$ can be expressed in the product form $P(\textbf{q}_{s}, \textbf{q}_{I})=P_s(\textbf{q}_{s}) P_I(\textbf{q}_{I})$. It can be checked from Eq. (\ref{inten-mom-gen}) that in this case no interference pattern will be observed and the information of the object will be absent in the photon counting rate measured by the camera. Therefore, in order the imaging scheme to work there must be some correlation between the momenta of the twin photons. Furthermore, the momentum correlation also determines the image quality. In fact, we will see in Sec. \ref{subsec:momcorr-img-reso} below that this momentum correlation limits the image resolution. 
\par
In the ideal scenario, when the momenta of the twin photons are perfectly correlated, the probability density $P(\textbf{q}_{s},\textbf{q}_{I})$ can be effectively replaced by a Dirac delta function. Consequently, it follows from Eq. (\ref{inten-mom-gen}) that
\begin{align} \label{inten-mom}
\mathcal{R}(\boldsymbol{\rho}_{c})\propto 1 + \left|T \left(\boldsymbol{\rho}_{o} \right) \right| \cos \big[\phi_{in} -\text{arg}\{T( \boldsymbol{\rho}_{o})\}\big],
\end{align}
The phase $\phi_{in}$ is varied experimentally and consequently an interference pattern is observed at each point $\boldsymbol{\rho}_{c}$ on the camera. It is evident from Eq. (\ref{inten-mom}) that the information of a point ($\boldsymbol{\rho}_{o}$) on the object appears in the interference pattern observed at a point ($\boldsymbol{\rho}_{c}$) on the camera. Extraction of this information results in imaging. 
\par
We illustrate the imaging by first considering an absorptive object for which we can set $\text{arg}\{T( \boldsymbol{\rho}_{o})\}=0$. It from Eqs. (\ref{vis-def}) and (\ref{inten-mom}) that the visibility of the single-photon interference pattern at a point $(\boldsymbol{\rho}_{c})$ on the camera is given by \cite{viswanathan2021position}
\begin{align} \label{abs-object-mom}
\mathcal{V}(\boldsymbol{\rho}_{c}) =\left|T(\boldsymbol{\rho}_{o})\right|. 
\end{align} 
Clearly, the spatially dependent visibility provides an image of the object. 
\par
\begin{figure}[ht]
\centering\includegraphics[width=\linewidth]{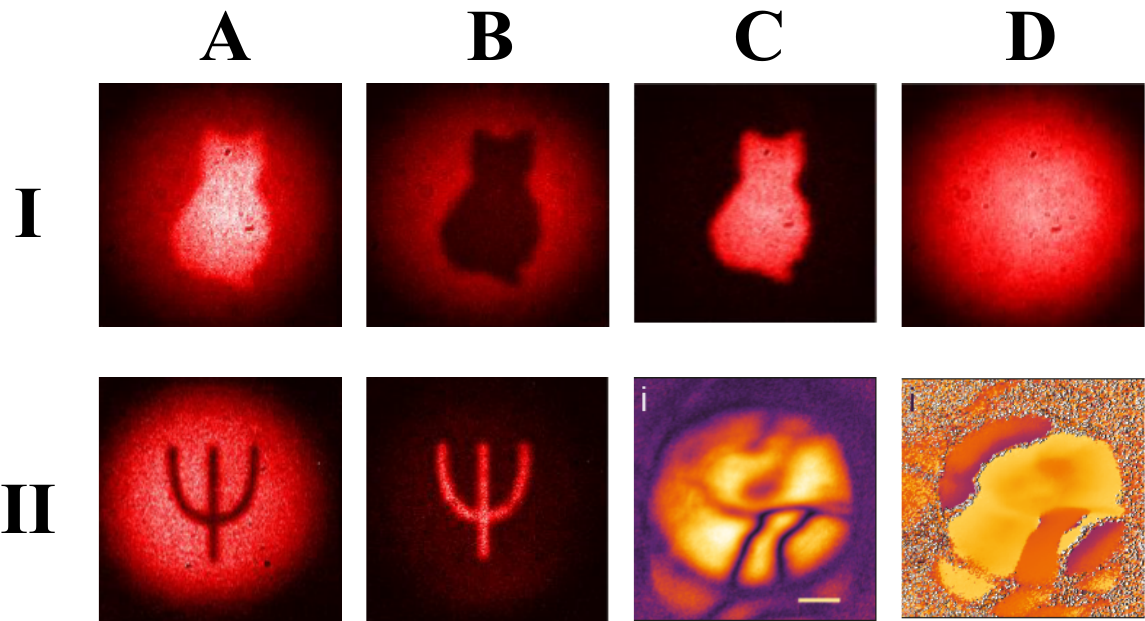}
\caption{\textbf{Absorption and phase imaging enabled by momentum correlations.} Figures I(A-D) and II(A-B) have been adapted from Ref.\cite{Lemos:2014p412}, which used the setup in Fig.\ref{fig:img-schm-mc}(a) and realized imaging enabled by momentum correlations. In \textbf{IA} and \textbf{IB} are shown two intensity signal outputs of a collinear non-degenerate ZWMI. The detection wavelength was $810\pm 1.5nm$, the sample was a cardboard cutout place at the Fourier plane of the sources and illuminated by an idler beam with wavelength centered at $1550 nm$. The difference (sum) of those two outputs is shown in  \textbf{IC (ID)}. Phase imaging of an etched silica plate using the same setup is shown in \textbf{IIA} and \textbf{IIB}. Momentum correlation enabled absorption (\textbf{IIC}) and phase (\textbf{IID}) images (adapted from Ref.\cite{sven-microscopy}) from a sample of a mouse heart. The setup was that shown in Fig.\ref{fig:herzog}(c) with the addition of lenses and an off-axis parabolic mirror. The detection and illumination central wavelengths were $0.8 \mu m$ and $3.8 \mu m$, respectively.}
\label{fig:Nature3}
\end{figure}

Alternatively, one can acquire the image of an absorptive object by subtracting the minimum intensity from the maximum intensity, \textit{i.e.}, by determining the quantity 
\begin{align} \label{image-fn}
G(\boldsymbol{\rho}_{c})=\mathcal{R}_{\text{max}}(\boldsymbol{\rho}_{c})-\mathcal{R}_{\text{min}}(\boldsymbol{\rho}_{c}).
\end{align}
In Ref. \cite{Lemos:2014p346}, images were acquired using this method. In Fig.\ref{fig:Nature3} IA and IB, interference is seen in the body of the cat, corresponding to regions of the idler field that are transmitted through a cardboard cutout. No interference is seen outside the cat, because the corresponding idler modes are blocked by the cardboard.  If one sums the two outputs the cat disappears (Fig.\ref{fig:Nature3}ID), and the Gaussian profile of the signal field is seen. This shows that the total signal field intensity is not affected by the absorptive object. Subtracting the two outputs results in a high contrast absorption image of the sample  (Fig.\ref{fig:Nature3}IC and IIC).

 We call $G(\boldsymbol{\rho}_{c})$ the \emph{image function}. It can be readily checked from Eq. (\ref{inten-mom}) that when the momenta of the twin photons are maximally correlated, $G(\boldsymbol{\rho}_{c}) \propto \left|T(\boldsymbol{\rho}_{o})\right|$.
\par
\par
Since $\text{arg}\{T( \boldsymbol{\rho}_{o})\}$ appears in Eq. (\ref{inten-mom}), phase imaging is also possible using this scheme (Fig.\ref{fig:Nature3} IIA, IIB and IID). For objects with relatively simple phase distribution as the ones considered in Ref. \cite{Lemos:2014p346}, the image can be obtained by intensity subtraction.  
\subsubsection{Image magnification}\label{subsec:momcorr-img-mag}
\begin{figure}[htbp]
	\centering
\includegraphics[width=0.7\linewidth]{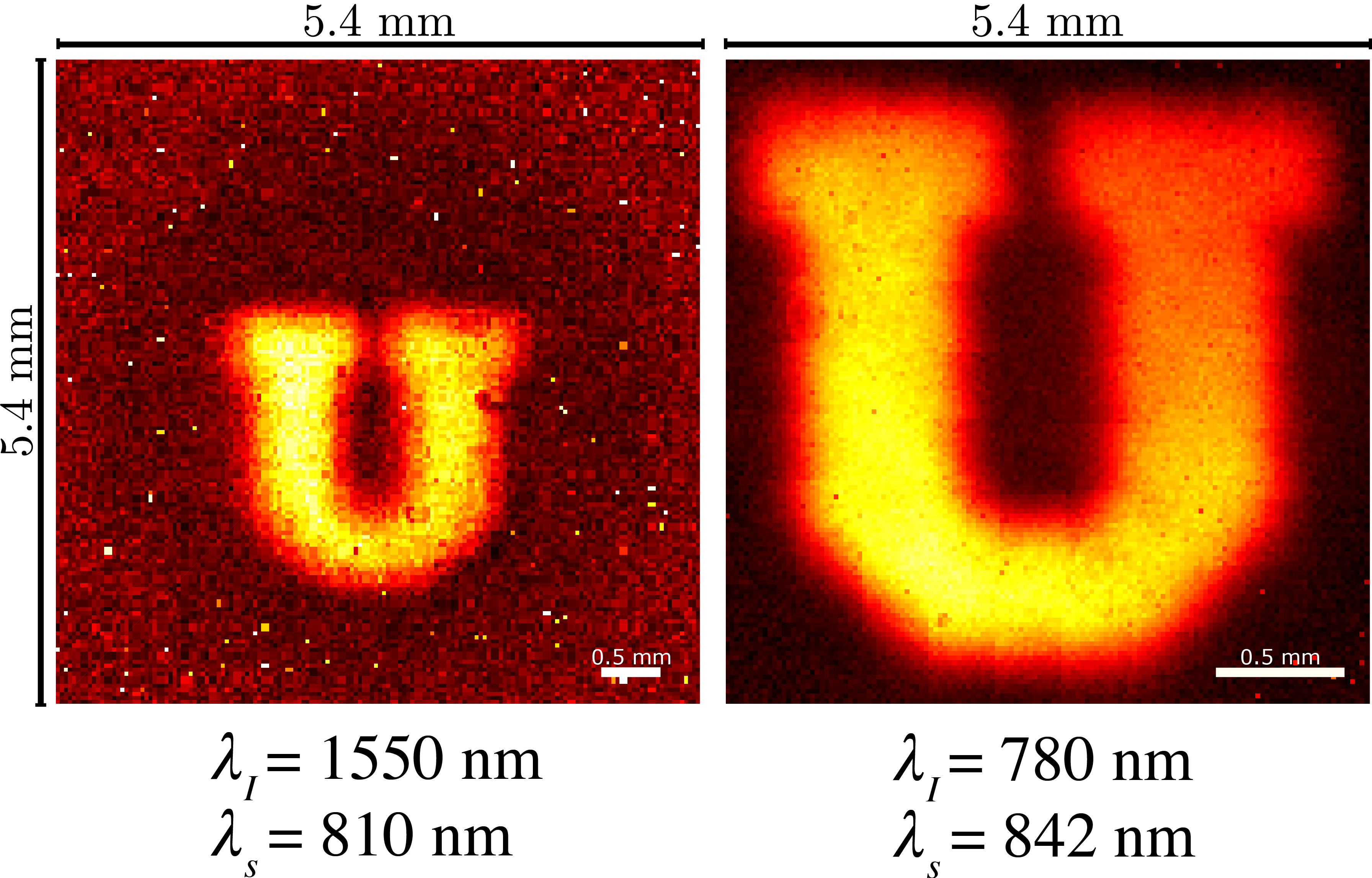}
	\caption{Image magnification in momentum correlation enabled QIUP. 
	The same object	is imaged for two sets of values of $\lambda_s$ and $\lambda_I$, while other parameters such as focal lengths and distances are unchanged. Higher value of the ratio $\lambda_s/\lambda_I$ resulted in larger image magnification (right). (Adapted from Fig. 2 of Ref. \cite{fuenzalida2020resolution}.)
    }
	\label{fig:magnification}
\end{figure}

It follows from Eqs. (\ref{inten-mom}) and (\ref{abs-object-mom}) that a point $\boldsymbol{\rho}_{o}$ on the object is imaged at a point $\boldsymbol{\rho}_{c}$ on the camera. Therefore, the image magnification ($M$) is equal to the ratio $|\boldsymbol{\rho}_{c}|/|\boldsymbol{\rho}_{o}|$. It now follows from Eqs. (\ref{q-obj-rel}) and (\ref{q-cam-rel}) that
\begin{equation} \label{image-mag-mom-1}
M \equiv \frac{|\boldsymbol{\rho}_{c}|}{|\boldsymbol{\rho}_{o}|} =\frac{f_c\lambda_{s}|\textbf{q}_s|}{f_I\lambda_{I}|\textbf{q}_I|}.
\end{equation} 
We now note that image blurring must be neglected for defining the magnification. Therefore, we must only consider the ideal case in which the momenta of the twin photons are perfectly correlated. As mentioned above, in this case the probability density governing the momentum correlation can be effectively replaced by a Dirac delta function. In fact, Eqs. (\ref{inten-mom}) and (\ref{abs-object-mom}) are obtained with this condition. In particular, we consider twin photons generated by SPDC, for which $P(\textbf{q}_{s}, \textbf{q}_{I}) \propto \delta(\textbf{q}_{s}+\textbf{q}_{I})$.  Consequently, for determining the image magnification, we need to use the condition $\textbf{q}_{s}=\textbf{q}_{I}$. Applying this condition to Eq. (\ref{image-mag-mom-1}), we find that the image magnification is given by
\begin{equation} \label{image-mag-mom}
M = \frac{f_c\lambda_{s}}{f_I\lambda_{I}}.
\end{equation} 
An interesting feature of QIUP in the far field configuration (i.e., enabled by momentum correlation) is that the image magnification depends on the wavelengths of twin photons. This fact is illustrated by Fig. \ref{fig:magnification}, which shows experimental observations presented in Ref. \cite{fuenzalida2020resolution}.
The wavelength dependence of magnification is also observed in the experiments reported in Refs. \cite{Lemos:2014p346,fuenzalida2020resolution,sven-microscopy,paterova2020hyperspectral}. 
\par
Thus far, we have not considered the sign of the magnification. This is because the sign will depend on the details of the lens systems using in the setup. For example, if the lens system is exactly as chosen in Refs. \cite{Lemos:2014p346}, the image magnification will have positive sign, i.e., the image will be erect. A detailed analysis of the image magnification for this case is presented in Ref. \cite{Lahiri2015a}.

\subsubsection{Spatial resolution}\label{subsec:momcorr-img-reso}
The resolution limit of momentum correlation enabled QIUP can be studied by applying theory discussed in Sec. \ref{subsec:momcorr-img-thoery}. A detailed description of this topic can be found in Ref. \cite{fuenzalida2020resolution}. Here, we discuss one resolution measure, namely the edge-spread function (ESF). We will consider another resolution measures in Sec. \ref{subsec:poscorr-img-reso} when we will discuss the resolution limit of the position correlation enabled QIUP. 
\par
We pointed out in Sec. \ref{subsec:momcorr-img-thoery} that the image of an absorptive object can be obtained by determining the image function, $G(\boldsymbol{\rho}_{c})$. It follows from Eqs. (\ref{inten-mom-gen}) and (\ref{image-fn}) that 
\begin{align} \label{image-MC}
G(\boldsymbol{\rho}_{c})\propto \int d\textbf{q}_{I} ~ P(\textbf{q}_{s}, \textbf{q}_{I}) \left|T (\boldsymbol{\rho}_{o}) \right|,
\end{align}
where $\boldsymbol{\rho}_{o}$ \& $\textbf{q}_{I}$ and $\boldsymbol{\rho}_{c}$ \& $\textbf{q}_{s}$ are related by Eqs. (\ref{q-obj-rel}) and (\ref{q-cam-rel}), respectively.
We will use the image function for determining the resolution because the mathematical analysis becomes simpler. We stress that the results remain the same if one obtains the image from the visibility.
\par
It is evident from Eq. (\ref{image-MC}) that when the momenta of the twin photons are not perfectly correlated, information about a range of points on the object plane appears at a single point on the camera. The broader the probability distribution $P(\textbf{q}_{s}, \textbf{q}_{I})$, the larger is the range of the points on the object plane. Therefore, it can be readily guessed that a weaker momentum correlation results in reduced resolution. 
\par
To study the resolution quantitatively, we need to know the form of the probability density function, $P(\textbf{q}_{s}, \textbf{q}_{I})$. To this end, we consider twin photons generated through SPDC and assume that the pump beam has a Gaussian profile. In this case, the probability density function can be approximated in the following form (see, for example, \cite{monken1998transfer,walborn10})
\begin{align}\label{prob-form-MC}
P(\textbf{q}_{s}, \textbf{q}_{I}) \propto  \exp\left(-\frac{1}{2}|\textbf{q}_{s} + \textbf{q}_{I}|^{2} w_{p}^{2} \right),
\end{align}
where $w_p$ represents the waist of the Gaussian pump beam. Clearly, the standard deviation of the probability distribution is inversely proportional to the pump waist. Consequently, a larger pump-waist ($w_p$) results in a narrower probability distribution, i.e., enhanced momentum correlation. 
\par
To determine the ESF a knife-edge can be used as an object. The image, which turns out to be a blurred edge, effectively represents the ESF. Without any loss of generality, we assume that the knife-edge is placed parallel to the $y_o$ axis and along the line $x_o=x_{0}^{\prime}$, such that the idler field is blocked form $x_o\leq x_0'$. Therefore, we can write 
\begin{align} \label{edge-obj}
T (\boldsymbol{\rho}_{o})\equiv T (x_o,y_o) = \left\{
\begin{array}{ll}
      0 & x_o \leq x_0', \\
      1 & x_o > x_0', \\
\end{array} 
\right. \quad \forall ~y_o.
\end{align}
It now follows from Eqs. (\ref{image-MC}), (\ref{prob-form-MC}), and (\ref{edge-obj}) that the ESF is given by 
\begin{align}\label{ESF-MC}
\text{ESF}(x_c)\propto G(\boldsymbol{\rho}_{c}) \propto \text{Erfc}\left( \frac{\sqrt{2}\pi w_p}{f_c \lambda_s} \left(x_c - M x_{0}^{\prime} \right) \right),
\end{align}
where Erfc is the complementary error function and $M$ is the wavelength dependent image magnification given by Eq. (\ref{image-mag-mom}). We stress that in order to determine the ESF one can also measure the position dependent visibility instead of the image function.
\begin{figure}[t]
	\centering
	\includegraphics[width=0.99\linewidth]{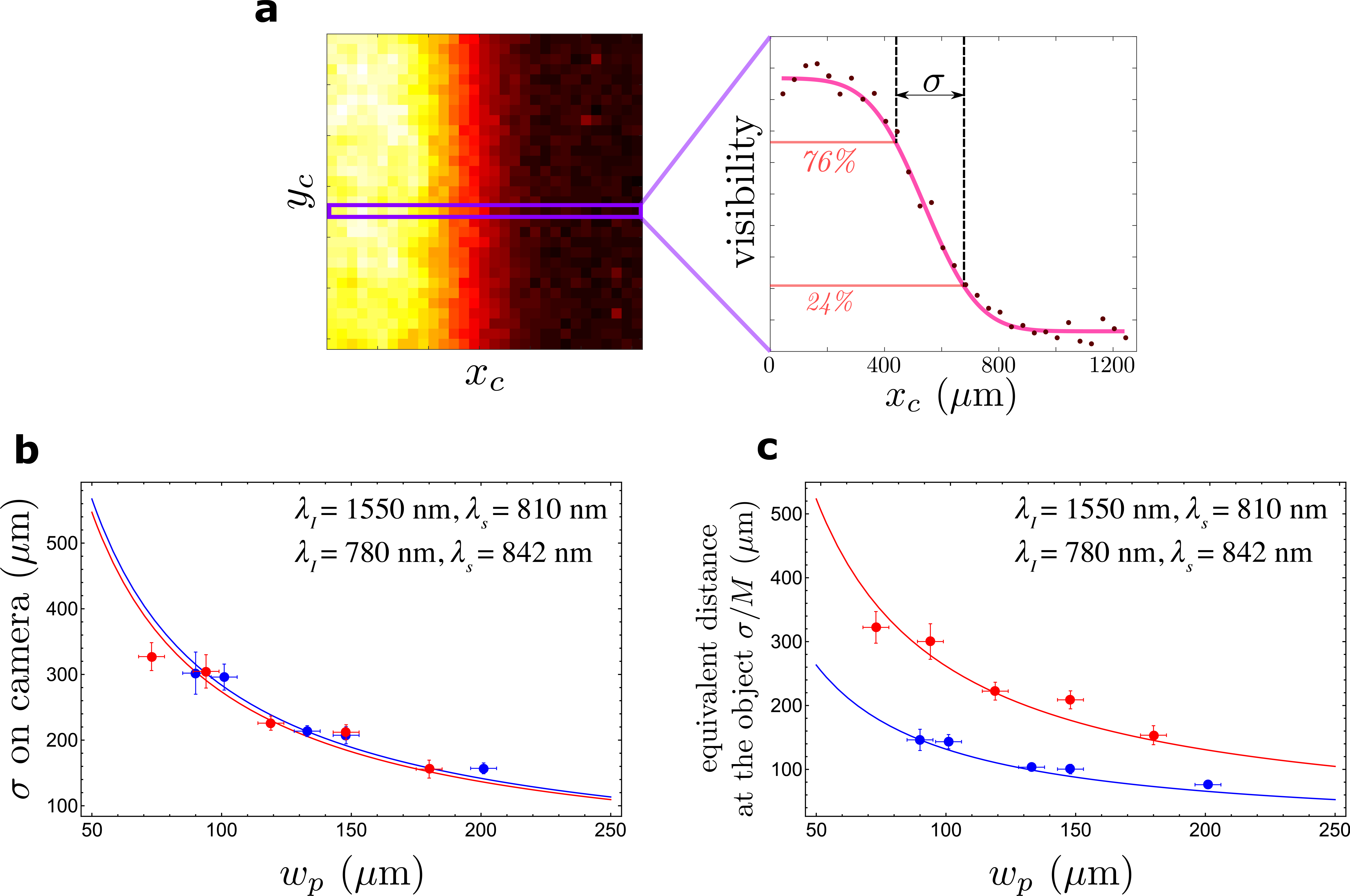}
\caption{Edge-spread function (ESF) and resolution. \textbf{a}, The image of a knife-edge is obtained by measuring the position dependent visibility on the camera (left). The visibility measured along an axis ($x_c$) is fitted with error function to experimentally determine the edge-spread function (right). The blurring ($\sigma$) is determined from the ESF. \textbf{b}, Experimentally measured values (data points) of $\sigma$ are compared with theoretical prediction (solid lines) for two sets of wavelengths, $\lambda_I=1550$ nm, $\lambda_s=810$ nm (red), and  $\lambda_I=780$ nm, $\lambda_s=842$ nm (blue). Since the detected wavelengths are close to each other, the blurring appears to be almost equal despite wide difference between the illuminating (undetected) wavelengths. \textbf{c}, The resolution ($\sigma/M$) is measured experimentally (data points) and compared with theoretical results (solid curves) for the same sets of wavelengths. Shorter illumination wavelength results in higher resolution. The resolution enhances with increasing pump-waist ($w_p$), i.e., with stronger momentum correlation between twin-photons. (Adapted from Fig. 4 of Ref. \cite{fuenzalida2020resolution})}
	\label{fig:esf-resultati}
\end{figure}
\par
Figure \ref{fig:esf-resultati}a shows an experimentally observer image of a knife-edge \cite{fuenzalida2020resolution}. The experimental results are in full accordance with the theoretical predictions made by Eq. (\ref{ESF-MC}). A measure of image blurring is how steeply the complementary error function representing the ESF rises. A sharper rise means less blurring. Mathematically, the blurring can be quantified by the inverse of the coefficient of $x_c$ inside the Erfc in Eq. (\ref{ESF-MC}), i.e., by
\begin{align}\label{spread-MC}
\sigma=\frac{f_c \lambda_s}{\sqrt{2}\pi w_p}.
\end{align}
This quantity can be determined from the experimentally obtained ESF (Fig. \ref{fig:esf-resultati}a): one can check from the properties of the complementary error function that $\sigma$ is the distance for which the value of ESF rises from $24\%$ to $76\%$ of the maximum attainable value. 
\par
\begin{figure}[htbp]
	\centering
	\includegraphics[width=0.99\linewidth]{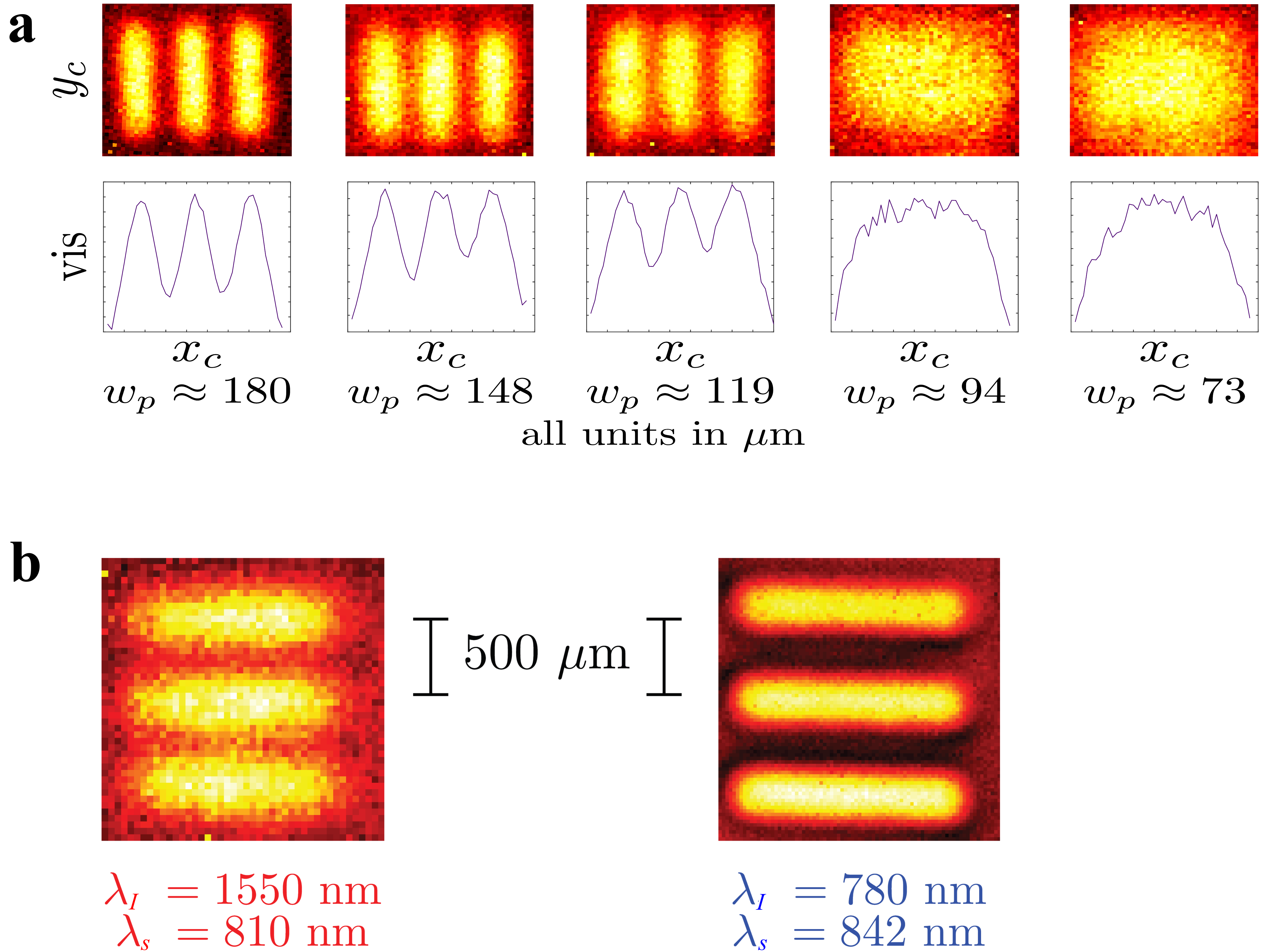}
	\caption{Resolution of momentum correlation enabled QIUP. \textbf{a}, Resolution enhances as momentum correlation becomes higher. A set of slits are imaged for five values of pump waist ($w_p$) in the decreasing order (left to right). A bigger value of $w_p$ means a stronger momentum correlation between the twin photons, which results in higher resolution. (Wavelengths are kept the same for each measurement.) \textbf{b}, A smaller value of the undetected wavelength ($\lambda_I$) results in better resolution. The same set of slits are imaged for $\lambda_I=1550$ nm (left) and $\lambda_I= 780$ nm (right), while the pump waist is kept the same. (Adapted from Figs.~3b and 5b of Ref.~\cite{fuenzalida2020resolution}.)}
	\label{fig:res-MC-summary}
\end{figure}

Equation (\ref{spread-MC}) shows how the resolution depends on the momentum correlation between the twin photons. As mentioned below Eq. (\ref{prob-form-MC}), a larger value of the pump waist ($w_p$) implies a stronger momentum correlation between the twin photons. It follows from Eq. (\ref{spread-MC}) that a larger value of $w_p$ results in a smaller value $\sigma$, i.e., less blurring implying higher resolution. Figure \ref{fig:esf-resultati}b shows the experimentally measured values of $\sigma$ for two experimental setups \cite{fuenzalida2020resolution}. The solid lines represent theoretical predictions made from Eq. (\ref{spread-MC}). Figure \ref{fig:res-MC-summary}a demonstrates how the image of a collection of three slits gets blurred when the momentum correlation between twin photons is reduced.
\par
We now discuss the wavelength dependence of the resolution. Equation (\ref{spread-MC}) shows that $\sigma$ does not depend on the wavelength ($\lambda_I$) of the undetected photon that interacts with the object; it instead depends on the wavelength ($\lambda_s$) of the detected photon that never interacts with the object. However, it must \emph{not} be concluded from this observation that the resolution depends on the detected wavelength.
This is because $\sigma$ is the blurring measured in the camera coordinates and the resolution is basically the minimum resolvable distance on the object place. Therefore, a more accurate measure of resolution is obtained if one divides $\sigma$ by the magnification, $M$, which essentially results in expressing $\sigma$ in object coordinates. The division by the magnification is essential for understanding the wavelength dependence of the resolution because in this imaging configuration the magnification depends on wavelength. It now follows from Eqs. (\ref{image-mag-mom}) and (\ref{spread-MC}) that 
\begin{align}\label{reso-MC}
{\rm res}=\frac{\sigma}{M}=\frac{f_I \lambda_I}{\sqrt{2}\pi w_p},
\end{align}
which is a measure of resolution of this imaging scheme. It follows from Eq. (\ref{reso-MC}) that the resolution depends only on the undetected wavelength, i.e., the wavelength that probes the object. Here, we note that the resolution depends on the momentum correlation in the same way $\sigma$ does. Therefore, our conclusions regarding the dependence of resolution on the momentum correlation remains unchanged. 
\par
The wavelength dependence of resolution has been verified experimentally by building two experimental setups for which the values of detected wavelength are very close (810 nm and 842 nm), whereas the undetected wavelengths are widely separated (1550 nm and 780 nm) \cite{fuenzalida2020resolution}. The experimental results are displayed in Fig. \ref{fig:esf-resultati}b, which confirms the prediction made by Eq. (\ref{spread-MC}): the values of $\sigma$ for the two setups are very close to each other because the values of detected wavelength are also very close. However, the experimentally obtained value of $\sigma/M$ shows that the resolution of the two setups are significantly different (Fig. \ref{fig:esf-resultati}c). This is because the undetected wavelength for the two setups are widely separated. The wavelength dependence of the resolution has also been experimentally tested by using a 1951 USAF resolution test chart and it has been found that the results matches accurately with theoretical predictions \cite{fuenzalida2020resolution}. In Fig. \ref{fig:res-MC-summary}b, we show images of the same collection of slits for two values of undetected wavelength while the pump waists (i.e., the momentum correlation) is kept fixed. It is evident from this figure that the resolution is higher for a shorter undetected wavelength.
\par
In refs.~\cite{sven-microscopy} and \cite{paterova2020hyperspectral}, the same imaging resolution Eq.\ref{reso-MC} was verified for the $SU(1,1)$ interferometer with illumination photons at the mid-infrared (~$3 \mu m$) and detection wavelength suitable for silicon based cameras. 
\par

\subsubsection{Resolution and Field-of-View}\label{subsec:reso-FoV-mc}

A useful parameter to calculate is the \textit{number of spatial modes per direction}, also sometimes referred to as "number of spatial modes" is typically estimated by the Field of View (FoV) divided by the spatial resolution.
The FoV is straight-forwardly given by the emission angle of the down-converted idler light that defines the size of the illuminating area:
\begin{equation}\label{eq.3}
{\rm FOV_{MC}}=2f_I\tan(\theta_I)\approx 2f_I\theta_I,
\end{equation}
where $f_I$ denotes the focal length of the collimating optical element adjacent to the crystal, $\theta_I$ is the idler divergence angle. Defined as half-width at half-maximum (HWHM) it is given by 
\begin{equation}
 \theta_I=\lambda_I\sqrt{\frac{2.78 n_s n_I}{ \pi L(n_s\lambda_I+n_I\lambda_s)}},
\end{equation}
where $n_s$ $ (n_I)$ is the index of refraction of the signal (idler) field in the crystal (see supplementary material of ref.\cite{sven-microscopy}).

The number of spatial modes per direction can therefore be estimated as:
\begin{equation}\label{eff-pixels}
m_{MC}=\frac{\rm FOV_{MC}}{{\rm res}^{\rm FWHM}}\propto w_p\sqrt{\frac{n}{L(\lambda_I+\lambda_s)}},
\end{equation}
where ${\rm res}^{\rm FWHM}=2\sqrt{ln2 }\;{\rm res}$ and  $n=n_s\approx n_I$. 
Unsurprisingly, the number of spatial modes per direction does not depend on $f_I$ or the magnification ($M$) - provided the intermediate optics features a sufficiently large numerical aperture.
\par
We conclude this section by summarizing the key features of the resolution of momentum correlation enabled QIUP. The resolution enhances if the momentum correlation between twin photon becomes stronger. For photon generated by SPDC, the resolution is linearly proportional to the standard deviation of the probability distribution that governs the momentum correlation. The resolution is linearly proportional to the wavelength of the undetected photon (i.e., the photon that probes the object). Therefore, a shorted undetected wavelength results in a higher correlation. 
\par
Finally, the method described in this section applies to any twin-photon state. Here, we considered a Gaussian probability distribution. For other forms of probability distribution, an exact mathematical expression may not be obtained and resorting to numerical simulation may be necessary.

\subsection{Imaging Enabled by Position Correlation}\label{subsec:poscorr-img}
\subsubsection{General theory}\label{subsec:poscorr-img-thoery}
We now analyze the complementary scenario in which both the object and the camera are placed in the near field relative to the source. In this case, the imaging is enabled by position correlation between the twin photons. We stress that we do not use evanescent fields for the image acquisition and therefore the our method is not to be confused with conventional near-field imaging. We simply place the object and the camera at separate image planes of the sources. 
\par
The joint probability density of detecting the signal and the idler photons at positions (transverse coordinates) $\boldsymbol{\rho}_{s}$ and $\boldsymbol{\rho}_{I}$, respectively, on the source plane is given by \cite{walborn10}
\begin{equation} \label{prob-pos}
P(\boldsymbol{\rho}_{s},\boldsymbol{\rho}_{I}) \propto \left|\int d \textbf{q}_{s} \ d \textbf{q}_{I}\  C(\textbf{q}_{s}, \textbf{q}_{I}) \ e^{i (\textbf{q}_{s}.\boldsymbol{\rho}_{s} + \textbf{q}_{I}.\boldsymbol{\rho}_{I})}\right|^{2},
\end{equation}
where $C(\textbf{q}_{s},\textbf{q}_{I})$ is introduced in Eq. (\ref{state-spdc}).
The position correlation between the two photons is governed by this joint probability density. If $P(\boldsymbol{\rho}_{s},\boldsymbol{\rho}_{I})$ can be expressed as a product of a function of $\boldsymbol{\rho}_{s}$ and a function of $\boldsymbol{\rho}_{I}$, there is no position correlation. In the other extreme case, when the positions of the two photons are maximally correlated, the joint probability density is proportional to a Dirac delta function. 
\par
The schematic of the imaging setup is given in Fig. \ref{fig:img-schm-mc}. In contrast to the case of momentum correlation enabled QIUP, in this case, the propagating part of the source field is recreated on the object by an imaging system.

As usual, beam $I_1$ from source $Q_1$ illuminates the object, passes through source $Q_2$, and gets perfectly aligned with beam $I_2$. An imaging system, $B$, is placed between the source $Q_1$ and the object ($O$) such that the idler field at $Q_1$ is imaged onto the object with magnification $M_I$. Another imaging system, $B'$, images the idler field at the object onto source $Q_2$ with magnification $1/M_I$ (i.e., demagnified by the equal amount). These two imaging systems also ensure that $Q_2$ lies on the image plane of $Q_1$. For simplicity, we have assumed that magnifications of $B$ and $B'$ have the same sign. In order to obtain the best possible alignment of beams $I_1$ and $I_2$, it is essential that the magnitude of the total magnification due to the combined effect of $B$ and $B'$ is 1. 
\par
The object is once again characterized by its complex amplitude transmission coefficient $T(\boldsymbol{\rho}_o)$, where $\boldsymbol{\rho}_o$ represents a point on the object plane. Since the object is treated like a beam splitter, the quantum field associated with the idler photon at $Q_2$ is related to that at $Q_1$ by the following formula \cite{viswanathan2021position}
\begin{align}\label{align-pos}
\hat{E}_{I_{2}}^{(+)}(\boldsymbol{\rho}_{I}) = e^{i \phi_{I}'(\boldsymbol{\rho}_{I})}\big[ e^{i \phi_{I}(\boldsymbol{\rho}_o)}  T (\boldsymbol{\rho}_o) \hat{E}_{I_{1}}^{(+)}(\boldsymbol{\rho}_{I})  +   R (\boldsymbol{\rho}_o) \hat{E}_{0}^{(+)}(\boldsymbol{\rho}_{I}) \big],
\end{align}
where $\hat{E}_{0}^{(+)}(\boldsymbol{\rho}_{I})$ is the corresponding vacuum field, $|T(\boldsymbol{\rho}_o)|^2+|R(\boldsymbol{\rho}_o)|^2=1$, phases $\phi_{I}(\boldsymbol{\rho}_{o})$ and $\phi_{I}'(\boldsymbol{\rho}_{I})$ are introduced by the imaging systems $B$ and $B'$, respectively, and due to the presence of these imaging systems, $\boldsymbol{\rho}_o=M_{I}$ and $\boldsymbol{\rho}_{I}$ are related by the formula
\begin{align}\label{obj-id-rel}
\boldsymbol{\rho}_o=M_{I} \boldsymbol{\rho}_{I}.
\end{align}
\par
It follows from Eq. (\ref{align-pos}) that (see \cite{viswanathan2021position} for a detailed proof)
\begin{align}\label{align-idler op}
\hat{a}_{I_{2}}(\textbf{q}_{I}) = \int &d \textbf{q}_{I}^{'} \frac{1}{M_{I}^2} \big[ \widetilde{T}'\left(\frac{\textbf{q}_{I}- \textbf{q}_{I}^{'}}{M_{I}} \right) \ \hat{a}_{I_{1}}(\textbf{q}_{I}^{'}) \nonumber \\
&+ \widetilde{R}'\left(\frac{\textbf{q}_{I}- \textbf{q}_{I}^{'} }{M_{I}} \right)  \ \hat{a}_{0}(\textbf{q}_{I}^{'}) \big],
\end{align}
where $\widetilde{T}'(\textbf{q}_{I}/M_{I})$ and $\widetilde{R}'(\textbf{q}_{I}/M_{I})$ are the Fourier transforms of $\text{exp}[i\{\phi_{I}(M_I\boldsymbol{\rho}_{I})+\phi_{I}'(\boldsymbol{\rho}_{I})\}] T(M_{I} \boldsymbol{\rho}_{I})$ and $\text{exp}[i\phi_{I}'(\boldsymbol{\rho}_{I})]R(M_{I} \boldsymbol{\rho}_{I})$, respectively. We encourage the readers to convince themselves that $\hat{a}_{0}$ is related to $\hat{E}_{0}^{(+)}$ in the same way $\hat{a}_{I_{j}}$ is related to $\hat{E}_{I_{j}}^{(+)}$.
\par
It becomes evident by comparing Eq. (\ref{align-idler op}) with Eq. (\ref{align-mom}) that the conditions due to the alignment of the idler beams are significantly different in near and far field QIUP. Although the initial quantum state generated by the two sources is once again given by Eq. (\ref{biphoton-state}), the difference between the alignment conditions ensures that the final quantum state of light generated by the imaging system is distinct in the two configurations. In the present scenario, the quantum state generated by the system is obtained by combining Eqs. (\ref{biphoton-state}) and (\ref{align-idler op}) and is given by  \cite{viswanathan2021position}
\begin{align} \label{biphoton-state-align}
|\psi\rangle =& \alpha_{1} \int d \textbf{q}_{I_{1}} \ d \textbf{q}_{s_{1}} \ C(\textbf{q}_{I_{1}}, \textbf{q}_{s_{1}}) \   |\textbf{q}_{I_{1}}\rangle_{I_{1}}  |\textbf{q}_{s_{1}}\rangle_{s_{1}} \nonumber \\
&+ \alpha_{2}  \int d \textbf{q}_{I_{2}}\ d \textbf{q}_{s_{2}} \ d \textbf{q}_{I}^{'}  \ C(\textbf{q}_{I_{2}}, \textbf{q}_{s_{2}}) \nonumber \\
& \qquad \times \frac{1}{M_{I}^2}\Big[\widetilde{T}'^{*}\left(\frac{\textbf{q}_{I_{2}}- \textbf{q}_{I}^{'}}{M_{I}} \right) \    |\textbf{q}^{'}_{I}\rangle_{I_{1}} \nonumber \\
&\qquad \qquad + \widetilde{R}'^{*}\left(\frac{\textbf{q}_{I_{2}}- \textbf{q}_{I}^{'} }{M_{I}} \right) \   |\textbf{q}^{'}_{I}\rangle_{0} \Big]  |\textbf{q}_{s_{2}}\rangle_{s_{2}},   
\end{align}
where $|\textbf{q}\rangle_{0}=\hat{a}_{0}^{\dag}(\textbf{q})|vac \rangle$. 
\par
The two signal beams ($S_{1}$ and $S_{2}$) are superposed by a $50:50$ beam splitter (BS) and one of the outputs of BS is detected by a camera. An imaging system ($A$) with magnification $M_S$ ensures that the signal field at the sources is imaged onto the camera (Fig. \ref{fig:img-schm-mc} a \& b). If we represent the signal field at each source ($z=0$) by its angular spectrum (\cite{MW}, Sec. 3.2; see also \cite{monken1998transfer}), the positive frequency part of the total signal field at a point, $\boldsymbol{\rho}_{c}$, on the camera is given by \cite{viswanathan2021position}
\begin{equation} \label{detection operator}
\hat{E}_{s}^{(+)}(\boldsymbol{\rho}_{c}) \propto \int d\textbf{q}_{s} \left[ \ \hat{a}_{s_{1}}(\textbf{q}_{s}) + i \ e^{[i \phi_{s0}+\phi_s(\boldsymbol{\rho}_{c})]} \ \hat{a}_{s_{2}}(\textbf{q}_{s}) \ \right] e^{i \textbf{q}_{s} \cdot \boldsymbol{\rho}_s},
\end{equation}
where the phase difference between the two signal fields is expressed as a sum of $\phi_{s0}$ and $\phi_s(\boldsymbol{\rho}_{c})$; the former is a spatially independent phase that can be varied to obtain interference patterns and the latter is a spatially dependent phase that can arise due to the presence of imaging system $A$. The presence of this imaging system results into the following relationship between the coordinates
\begin{align}\label{cam-sig-rel}
\boldsymbol{\rho}_{c} = M_{s}\boldsymbol{\rho}_{s}.
\end{align}
\par
The photon counting rate at a point $\boldsymbol{\rho}_{c}$ on the camera is determined by the standard formula $\mathcal{R}(\boldsymbol{\rho}_{c}) \propto \langle\psi|\hat{E}_{s}^{(-)}(\boldsymbol{\rho}_{c}) \hat{E}_{s}^{(+)}(\boldsymbol{\rho}_{c})|\psi\rangle,$
where $ \hat{E}_{s}^{(-)}(\boldsymbol{\rho}_{c}) = [\hat{E}_{s}^{(+)}(\boldsymbol{\rho}_{c})]^{\dagger}$. Using Eqs.  (\ref{prob-pos}), (\ref{obj-id-rel}), and (\ref{biphoton-state-align}) -- (\ref{cam-sig-rel}), we find that \cite{viswanathan2021position}
\begin{align} \label{photon count-final-expr}
\mathcal{R}(\boldsymbol{\rho}_{c}) \propto &  \int d \boldsymbol{\rho}_{I} P(\boldsymbol{\rho}_s,\boldsymbol{\rho}_I) \big( 1 + |T(\boldsymbol{\rho}_{o})|   \nonumber \\
& \quad \times \text{cos}[\phi_{in}+\phi_s(\boldsymbol{\rho}_{c}) -\phi_{I}(\boldsymbol{\rho}_{o})-\phi_{I}'(\frac{\boldsymbol{\rho}_{o}}{M_I})- \phi_{T}(\boldsymbol{\rho}_{o}) ]\big),
\end{align} 
where $\phi_{in}=\phi_{s0} + \text{arg}\{\alpha_{2}\} - \text{arg}\{\alpha_{1}\}$ and we have assumed $|\alpha_{1}|=|\alpha_{2}| = 1/\sqrt{2}$ for simplicity. 
\par
It follows from Eq. (\ref{photon count-final-expr}) that the information about the object (both magnitude and phase of the amplitude transmission coefficient) appears in the interference pattern observed on the camera, even though the photons probing with the object are not detected by the camera. The presence of $P(\boldsymbol{\rho}_s,\boldsymbol{\rho}_I)$ in Eq. (\ref{photon count-final-expr}) shows that the position correlation between the twin photons enables the image acquisition. For example, when there is no correlation between the momenta, $P(\boldsymbol{\rho}_s,\boldsymbol{\rho}_I)$ can be expressed in the product form $P(\boldsymbol{\rho}_s,\boldsymbol{\rho}_I)=P_s(\boldsymbol{\rho}_{s}) P_I(\boldsymbol{\rho}_{I})$. It can be checked from Eq. (\ref{photon count-final-expr}) that, in this case, no interference pattern will be observed and the information of the object will be absent in the photon counting rate measured by the camera. Therefore, in order the imaging scheme to work there must be some correlation between the positions of the twin photons. Furthermore, the position correlation also determines the image resolution; we will elaborate on this in Sec. \ref{subsec:poscorr-img-reso}. 
\par
In order to illustrate the image formation, we consider the special case in which the positions of the photon pair are maximally correlated, i.e., $P(\boldsymbol{\rho}_{s}, \boldsymbol{\rho}_{I}) \propto \delta(\boldsymbol{\rho}_{s}- \boldsymbol{\rho}_{I})$. We now have from Eq. (\ref{photon count-final-expr}) that \cite{viswanathan2021position}
\begin{align} \label{Photon count-delta}
&\mathcal{R}(\boldsymbol{\rho}_{c})\propto 1 + \left|T \left(\boldsymbol{\rho}_{o} \right) \right| \cos \big[\phi_{in} - \phi_{T}\left( \boldsymbol{\rho}_{o} \big)\right],
\end{align}
where, for simplicity, we have assumed that the phases introduced by the imaging systems are spatially independent and have included them in $\phi_{in}$. It is evident that if $\phi_{in}$ is varied, the photon counting rate (intensity) at each point on the camera varies sinusoidally, i.e., a single-photon interference pattern is observed at each point on the camera. 
\par
The image is acquired from these interference patterns in the same way as discussed in Sec. \ref{subsec:momcorr-img-thoery}. For example, when the object is purely absorptive, i.e., when $T(\boldsymbol{\rho}_{o})=|T(\boldsymbol{\rho}_{o})|$, the image is given by the spatially dependent visibility measured on the camera \cite{viswanathan2021position}:
\begin{align} \label{abs.object}
\mathcal{V}(\boldsymbol{\rho}_{c}) =\left|T(\boldsymbol{\rho}_{o})\right|.
\end{align}
Alternatively, the image can also be obtained by measuring the image function that is given by Eq. (\ref{image-fn}). It can be verified from Eq. (\ref{photon count-final-expr}) that both the visibility and the image function are independent of the phases introduced by the imaging systems. Therefore, the image of an absorptive object can be acquired using this method even if the phases introduced by the imaging systems are spatially dependent. However, image acquisition of a phase object may require information about these phases. 

\subsubsection{Image magnification}\label{subsec:poscorr-img-mag}
In order to determine the magnification, we once again need to neglect image blurring. Therefore, we must consider the case in which the twin photons are maximally position-correlated, i.e., $P(\boldsymbol{\rho}_{s}, \boldsymbol{\rho}_{I}) \propto \delta(\boldsymbol{\rho}_{s}- \boldsymbol{\rho}_{I})$. It now follows from Eqs. (\ref{obj-id-rel}) and (\ref{cam-sig-rel}) that the image magnification is given by \cite{viswanathan2021position}
\begin{equation} \label{image-mag-2}
M = \frac{M_{s}}{M_{I}}.
\end{equation} 
Clearly, the magnification does not have any explicit dependence on the wavelengths of the photons. This fact marks one important distinction between the momentum correlation enabled and position correlation enabled imaging.  

\subsubsection{Spatial resolution}\label{subsec:poscorr-img-reso}
We note from Eq. (\ref{photon count-final-expr}) that information about a range of points on the object plane, averaged by the joint probability density $P(\boldsymbol{\rho}_s,\boldsymbol{\rho}_I)$, appears at a single point on the camera. Since this probability distribution characterized the position correlation between the twin photons (see Eq. (\ref{prob-pos})), it becomes evident that the position correlation plays the key role in the image formation in this case.
\par
The resolution of position correlation enabled QIUP will be studied following the method described in Sec. \ref{subsec:momcorr-img-reso}. However, we choose a different resolution measure, namely the minimum resolvable distance. In this case, we consider two radially opposite points, separated by a distance $d$ and located on an axis (say, $X_o$) on the object plane. These two points can be represented by the amplitude transmission coefficient
\begin{align} \label{two radial points}
T(\boldsymbol{\rho}_{o})\equiv T(x_o,y_o) \propto \delta(y_{o})[\delta(x_{o}-d/2) + \delta(x_{o} + d/2)], 
\end{align}
where $x_{o}$ and $y_{o}$ represent the position along two mutually orthogonal Cartesian coordinate axes $X_o$ and $Y_o$, respectively.
\par
The image of the pair of points can be obtained by determining the image function. It follows from Eqs. (\ref{image-fn}) and (\ref{photon count-final-expr}) that for the position correlation enabled QIUP, the image function takes the form
\begin{align} \label{image-PC}
G(\boldsymbol{\rho}_{c})\propto \int d\boldsymbol{\rho}_{o}
P\left(\frac{\boldsymbol{\rho}_{c}}{M_{s}}, \frac{\boldsymbol{\rho}_{o}}{M_{I}} \right) |T(\boldsymbol{\rho}_{o})|.
\end{align}
\par
To assume a form of the joint probability density, we once again assume that the twin photons are generated through SPDC. In this case, one can represent the probability density function in the following form \cite{viswanathan2021resolution}:
\begin{align} \label{Prob Gaussian}
P \left( \frac{\boldsymbol{\rho}_{c}}{M_{s}}, \frac{\boldsymbol{\rho}_{o}}{M_{I}}\right) \propto \exp \left[-\frac{4 \pi}{L(\lambda_{I} + \lambda_{s})} \left|\frac{\boldsymbol{\rho}_{c}}{M_{s}} - \frac{\boldsymbol{\rho}_{o}}{M_{I}} \right|^{2} \right],
\end{align}
where $L$ represents the length of the nonlinear crystal. The standard deviation of this probability distribution is linearly proportional to $\sqrt{L}$. Therefore, a shorter crystal generates stronger position correlation between twin photons. 
\par
\begin{figure}[t]
\centering 
\includegraphics[width=0.99\linewidth]{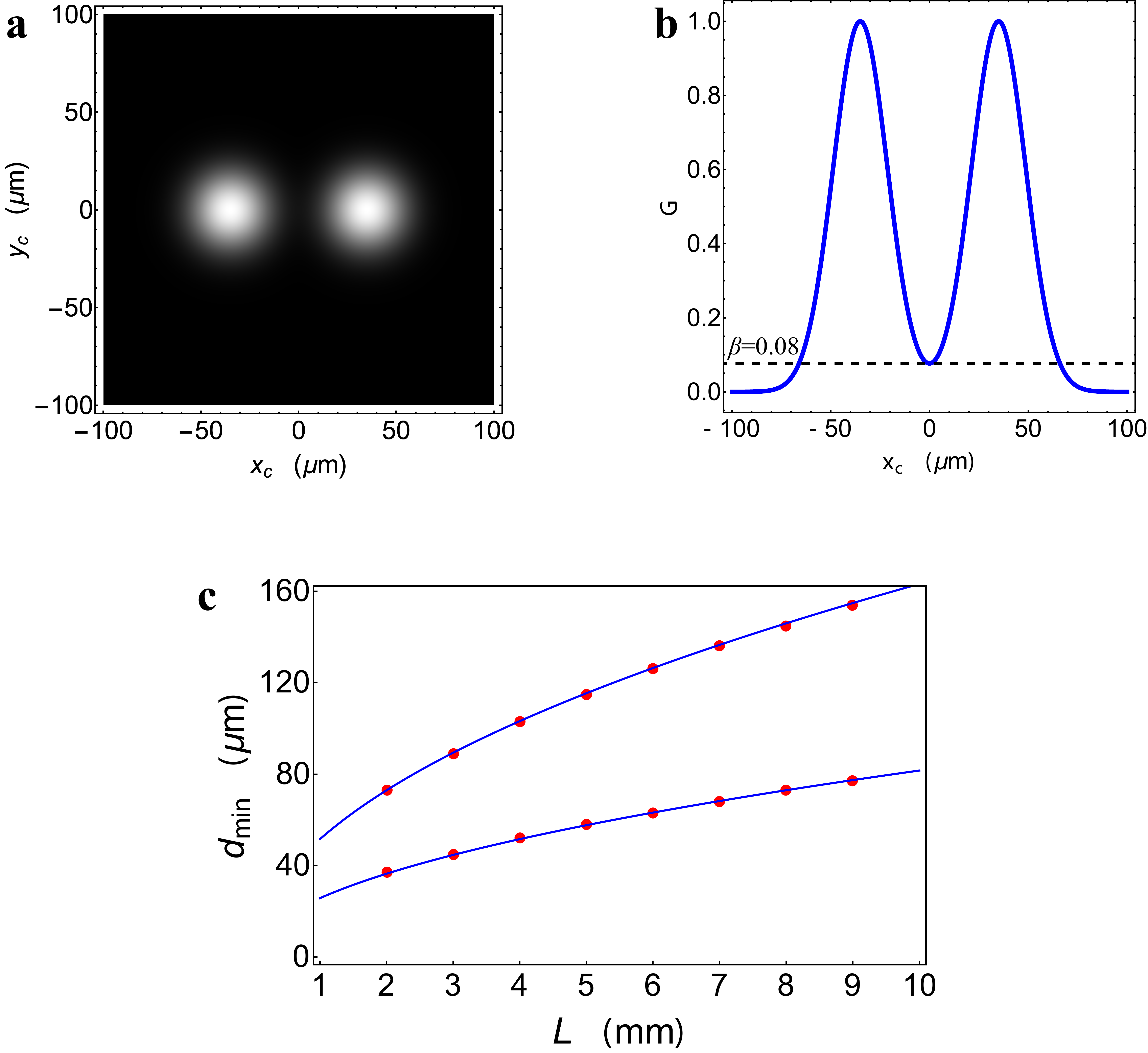}
\caption{Resolution and position correlation between twin photons. \textbf{a}, Simulated camera image of two points separated by a distance of $d=70$ $\mu$m for the following choice of parameters: $L=2$ mm, $\lambda_{s} = 810$ nm, $\lambda_{I} = 1550$ nm, and $M_{s} = M_{I} = 1$. \textbf{b}, The image function, $G(x_{c},0)$, plotted against $x_c$ the same set of parameters. The ratio ($\beta$) of its value at the dip to that at one of the peaks is $\beta\approx 0.08$. \textbf{c}, The minimum resolvable distance ($d_\text{min}$) plotted against crystal length ($L$) for $M_I=1$ and $M_I=2$ using Eq. (\ref{mindist}) (solid lines). The filled circles represent simulated data points for a pair of square pinholes with side length $1$ $\mu$m. The minimum resolvable distance increases (i.e., resolution reduces) as the position correlation becomes weaker. The resolution also decreases as the imaging magnification, $M_I$, from the source to the object increases. (Remaining parameters are same as in a and b.) (Adapted from Figs. 3c, 3d, and 4c of Ref. \cite{viswanathan2021resolution}.)}
\label{fig:VisbPlot}
\end{figure}

Using Eqs. (\ref{two radial points}) -- (\ref{Prob Gaussian}), we find that the image of the two points is given by the image function
\begin{align} \label{Res radial points}
G(\boldsymbol{\rho}_{c}) \propto & \exp\left[-\frac{4 \pi y_c^2}{M_s^2L(\lambda_{I} + \lambda_{s})} \right] \nonumber \\ &\times \Bigg\{\exp\left[-\frac{4 \pi}{L(\lambda_{I} + \lambda_{s})} \left(\frac{x_{c}}{M_{s}} -\frac{d}{2 M_{I}} \right)^{2} \right] \nonumber \\ & \qquad + \exp\left[-\frac{4 \pi}{L(\lambda_{I} + \lambda_{s})} \left(\frac{x_{c}}{M_{s}} +\frac{d}{2 M_{I}} \right)^{2} \right]\Bigg\}.
\end{align}
Figure \ref{fig:VisbPlot}a displays a simulated image of pair of points separated by $70$ $\mu$m for $L=2$ mm. The other parameters are given in the figure caption. 
In order to quantify how resolved the two points are we consider the function $G(x_{c},0)$, which is obtained by setting $y_c=0$ in the image function. If we plot $G(x_{c},0)$ against $x_c$, we get a double-humped curve which is illustrated by Fig. \ref{fig:VisbPlot}b. A measure of how well two points are resolved can be given by the ratio ($\beta$) of the value of $G$ at the dip ($G_{\text{dip}}$) to that at one of the peaks ($G_{\text{peak}}$), i.e., 
\begin{align}\label{ratio}
\beta\equiv \frac{G_\text{dip}}{G_\text{peak}}.
\end{align}
The lower the value of $\beta$, the better resolved the two points are. 
\par
The two points cannot be resolved after $\beta$ exceeds a certain value, say, $\beta_\text{max}$. We say that the two points are just resolved when $\beta=\beta_\text{max}$; in this case, the separation between the two points becomes the minimum resolvable distance (i.e., $d=d_\text{min}$). It follows from Eqs. (\ref{Res radial points}) and (\ref{ratio}) that $d_\text{min}\propto M_I \sqrt{L (\lambda_{I} + \lambda_{s})}$. (A detailed explanation is provided in Ref. \cite{viswanathan2021resolution}.)
In order to obtain a specific value of $d_\text{min}$, one needs to specify the value of $\beta$. There is no strict rule to choose the value of $\beta_\text{max}$. For the purpose of illustration,
we choose $\beta_\text{max}=0.81$ which appears in the study of fine structure of the spectral lines with a Fabry-Perot interferometer (\cite{BW}, Sec. 7.6.3). In this case, we numerically obtain the value of the proportionality constant and find it to be approximately 0.53. That is, the minimum resolvable distance defined by setting $\beta_\text{max}=0.81$ is given by the formula
\begin{align} \label{mindist}
d_{\text{min}} \approx 0.53 M_I \sqrt{L (\lambda_{I} + \lambda_{s})},
\end{align}
\par
To test the formula for minimum resolvable distance, we consider a pair of identical square apertures, each with side length $1$ $\mu$m, placed radially opposite on the $X_o$ axis (object plane). We choose nine values of the crystal length ($L$) and for each crystal length (i.e., fixed amount of position correlation), we choose two values of $M_I$. In each case, we numerically simulate the distance between the centers of the apertures by setting $\beta_\text{max}=0.81$. In Fig. \ref{fig:VisbPlot}c, we compare these numerically simulated distances (data points represented by filled circles) with theoretically predicted minimum resolvable distances (solid curves) that are predicted by Eq. (\ref{mindist}). Clearly, the simulated data are in excellent agreement with the theoretical prediction. 
\par
In Ref.~\cite{kviatkovsky2021} the same resolution was found by employing an SU(1,1) interferometer. Instead of the minimum resolvable distance [see Eq.~(\ref{mindist})] they define the resolution through the corresponding full width half maximum of the edge spread function on the camera ($\sigma^{\rm FWHM}_{PC}$): 
\begin{equation}\label{sigma-pc}
{\rm res}_{\rm PC}^{\rm FWHM}=\frac{\sigma_{\rm PC}^{\rm FWHM}}{M}=0.44 M_I \sqrt{\frac{ L(\lambda_I+\lambda_s)}{n} },   
\end{equation}
\par
where we assumed the same index of refraction, $n$, for signal and idler fields in the crystal. 
Equations (\ref{mindist}) and (\ref{sigma-pc}) reveal important features of the resolution limit of the position correlation enabled QIUP. We find that the resolution linearly proportional to the square root of the crystal length. Since a shorter crystal length implies a stronger position correlation between the twin photons, it becomes evident that a stronger position correlation between the twin photons results in a higher spatial resolution.
\par
Furthermore, the resolution is also linearly proportional to the magnification ($M_I$) of the imaging system, $B$, placed on the path of the undetected photon. Therefore, if the cross-section of the undetected beam (at source) demagnified while illuminating the object, the spatial resolution enhances, i.e., the resolution can be enhanced at the cost of the field of view (FoV). It is to be noted that a smaller value of $M_I$ results in a higher magnification of the imaging system (see Eq. (\ref{image-mag-2})). Therefore, if the image magnification is enhanced by using the optical components placed on the path of the undetected photon, the resolution also enhances. However, if one enhances the magnification using the optical elements placed in the path of the detected photon, resolution does not change.

\subsubsection{Resolution and Field of View}\label{subsec:reso-FoV-pc}
For a Gaussian pump beam, the Field of View (FoV) is a Gaussian distribution with a full width at half maximum (FWHM) given by \cite{kviatkovsky2021}
\begin{equation}\label{FoV}
FOV_{PC}=\sqrt{2ln2}M_Iw_p.    
\end{equation}
The ratio of the FoV and ${\rm res}_{\rm PC}^{\rm FWHM}$ approximates the number of spatial modes per direction,
\begin{equation}\label{s.modes}
m_{PC}=\frac{FOV_{PC}}{{\rm res}_{\rm PC}^{\rm FWHM}}\propto w_p\sqrt{\frac{n}{L(\lambda_I+\lambda_s)}}.
\end{equation}
We conclude this section by summarizing the main results relating to the resolution of position correlation enabled QIUP. The resolution enhances if the position correlation between twin photon becomes stronger. For photon generated by SPDC, the resolution is linearly proportional to the square root of the crystal-length, which is linearly proportional to standard deviation of the probability distribution corresponding to the position correlation. Both the detected and undetected wavelengths play symmetric role in determining the resolution and the resolution can be enhanced at the cost of the field of view. 
\par
The method described in this section applies to any twin-photon state with in idealized angular distribution such as the here considered Gaussian probability distribution. For other forms of probability distribution (see, for example, \cite{reichert2017quality}), an exact mathematical expression may not be obtained and resorting to numerical simulation may be necessary.

\subsection{Comparison between momentum and position correlations enabled QIUP}\label{subsec:comp}
Equations (\ref{eff-pixels}) and (\ref{s.modes}) show that the number of spatial modes per direction has the same dependence on experimental parameters in position correlation and in momentum correlation enabled QUIP. 
However, these results were obtained considering the Gaussian approximation for the $sinc^2$-shaped angular emission probability, which for the case of standard collinear SPDC leads to significant deviations in the exact number of spatial modes per direction\cite{kviatkovsky2021}. This is particularly noteworthy in the case of position correlation enabled imaging, where the number of spatial modes per direction is significantly reduced, compared to imaging via momentum correlation\cite{kviatkovsky2021}.
\par
In Table \ref{table:comp}, we summarize the comparison between QIUP schemes that utilize momentum correlation and position correlation between twin photons.
\begin{table*}[t]
\centering
\label{T:equipos}
\begin{center}
\begin{tabular}{| c | c | c | c | c | c |}
\hline
\cellcolor{black!25} & \multicolumn{3}{ c |}{\textbf{Imaging with Momentum Correlations}} &\multicolumn{2}{ c |}{\textbf{Imaging with Position Correlations}} \\ 
\cline{2-6}
\cellcolor{black!25} & \thead{\shortstack{\textbf{Formula}\\ (cf. \cite{fuenzalida2020resolution})}} & \thead{\shortstack{\textbf{Experiment (Theory)}\\ (cf. \cite{fuenzalida2020resolution})}} & \thead{\shortstack{\textbf{Experiment (Theory)}\\ (cf. \cite{sven-microscopy})}} & \thead{\shortstack{\textbf{Formula}\\ (cf. \cite{viswanathan2021resolution})}} & \thead{\shortstack{\textbf{Experiment (Theory)}\\ (cf. \cite{kviatkovsky2021})}} \\
\hline
res$^{\rm FWHM}$ &$\frac{\sqrt{2 ln2} f_I \lambda_I }{\pi w_p}$& $366~\mu m ~(366~\mu \text{m})$ &$320~\mu \text{m} ~(330~\mu \text{m})$ &$0.44M_I \sqrt{\frac{ L(\lambda_I+\lambda_s)}{n}}$ & $9~\mu \text{m}~(8~\mu \text{m})$\\ \hline
FOV& $2f_I \theta_i$ &--- ~~(3.7~\text{mm})& $9~\text{mm}~ (10~\text{mm})$ & $\sqrt{2 ln 2} w_p M_I$ & $160~\mu \text{m}~ (130~\mu \text{m})$ \\ \hline
$m$ & $5w_p\sqrt{\frac{ n}{L(\lambda_I+\lambda_s)}}$ & --- ~~ (10) & $28~(30)$ & $2.7w_p\sqrt{\frac{n}{ L(\lambda_I+\lambda_s)}}$ & $18~(16)$ \\ \hline
\end{tabular}
\caption{\textbf{Experimental and theoretical comparison between QUIP enabled by momentum and position correlations.} Experimental parameters for QIUP enabled by momentum correlations in ref.\cite{fuenzalida2020resolution}: $\lambda_I=1550nm,\lambda_s=810 nm, n_s\approx n_I=1.4, f_I=75, w_p= 119 \mu m$, experimentally measured  res= res$^{FWHM}/(2\sqrt{ln2})=220 \mu m. $ Experimental parameters for QIUP enabled by momentum correlations in ref.\cite{sven-microscopy}:
L=2mm, $n_s\approx n_I=1.8$, $w_p$= 430$\mu m$, $M_I=1$,    $f_I$=100mm, $\lambda_I$=3.8$\mu m$, $\lambda_s$=0.8, $\theta_i=50mrad$. Experimental parameters for QUIP enabled by position correlations in ref.\cite{kviatkovsky2021}: L=2mm, $n_s\approx n_I=1.8$, $w_p$= 430$\mu m$,$M_I=0.25$, $\lambda_I$=3.8$\mu m$, $\lambda_s$=0.8,$\theta_i=50 m$rad.  }\label{table:comp}
\end{center}
\end{table*}

\section{Further Applications} \label{sec:applications}
A main theme of applications for nonlinear interferometers is to enable measurements with high-performance Silicon-based cameras or linecameras (arrays) for wavelength regions, where these and also corresponding sources are not easily available. These are mainly regions in the infra-red. The applications can be divided into using spatial or spectral correlations  or both. 

Using spatial correlations enables imaging techniques such as phase-imaging, amplitude-imaging, as we have shown in the previous section, which can also can be extended to microscopy and holography, which is a combination of phase and amplitude imaging. Spectral correlations enable spectroscopy and optical coherence tomography (OCT). Applying nonlinear interferometers to hyperspectral imaging represents the combination of using spatial and spectral correlations. Note that for spectral measurements using nonlinear interferomenters based on SPDC one can provide very large spectral bandwidths with specially designed crystals\cite{Vanselow2019}. Such large bandwidths are typically otherwise complex and costly to achieve.

Also applications accessing the THz-region are possible, addressing mainly OCT-type measurements\cite{kutas2020} with a single spatial mode, since multi-spatial mode correlation is extremely challenging to achieve (see "Limitations").

Another special class of applications is refractometry for not easily accessible wavelength regions\cite{Paterova2018}. Here the reflectivity and/or field transmittance of a dielectric medium can be measured (wavelength dependently), allowing to infer its (wavelength-dependent) refractive index.

\subsection{Holography and 3D imaging}
The first quantitative phase imaging with undetected photons,i.e. extraction of the value of phase at each pixel of the image,  was realized in ref.\cite{fuenzalida2020resolution} and was later realized in ref.\cite{topfer2022quantum}. Simple  subtraction of the the outputs is not enough to realize quantitative phase imaging. Fortunately, the interferogram described by Eqs \ref{inten-mom} and \ref{Photon count-delta} has a form very similar to standard interferograms, and so the well-known digital holography methods can be used to extract the phase of the object $\text{arg}\{T( \boldsymbol{\rho}_{o})\}$. 

The phase stepping method requires recording at least three interference patterns with different values of $\phi_{in}$ given by $\phi_{in}^j=\frac{2\pi j}{K}$, where K is the number of phase steps. The interferograms obtained for different phases need to be added with appropriately chosen complex prefactors. The resultant complex-valued function of position contains the phase information as its argument. 

Disadvantages of phase stepping include the necessity of precisely controlling $\phi_{in}$ and recording multiple images from which a single phase image is reconstructed. The sample and the setup must not change during the acquisition of the images.

The Fourier off-axis holography method dispenses with the need of acquiring multiple frames at a cost of slightly more complex image processing and a small modification to the experimental setup. Here an extra tilt, or equivalently a linear phase $\phi_{in}$, is introduced between the interfering beams at the object plane. This linear phase $\phi_{in}(\rho_0)=a \rho_0$ enables the isolation of the second term in Eqs \ref{inten-mom} and \ref{Photon count-delta} , from which the phase can be extracted by taking its argument. The isolation of the phase is performed by filtering the interferogram in the Fourier domain. The linear phase has to be subtracted from the reconstructed phase and therefore it has to be pre-calibrated.

Interestingly, digital holography can enable three-dimensional ($3$D) imaging. Multiple phase images, taken at different illumination angles can be combined algorithmically to obtain a single $3$D image \cite{Field_tomography}.

\subsection{Optical Coherence Tomography}\label{subsec:OCT}
\begin{figure}[htbp]
\centering\includegraphics[width=\linewidth]{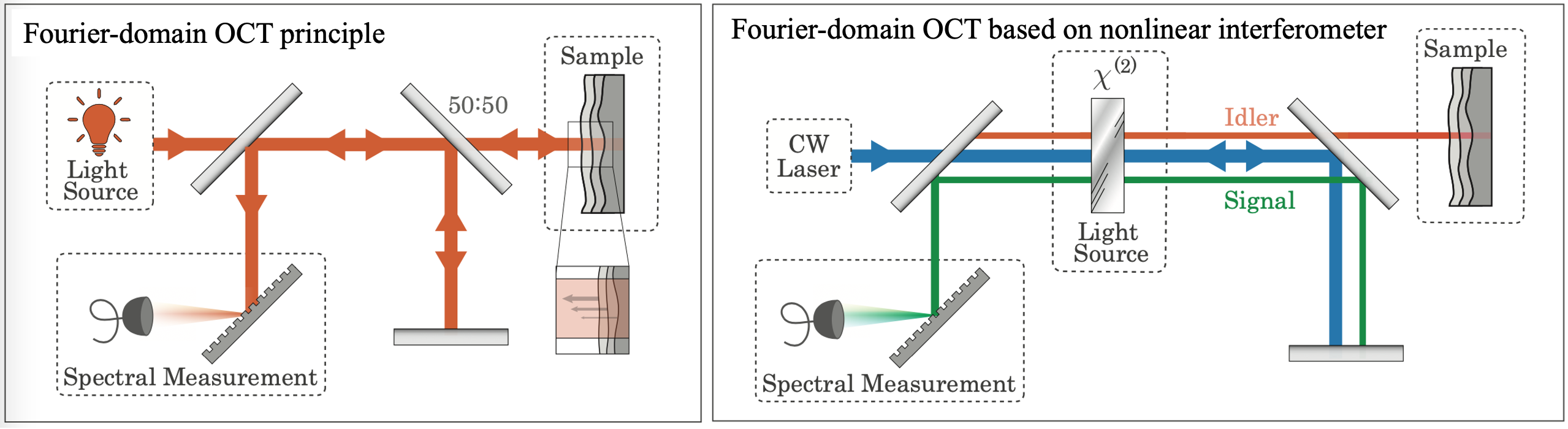}
\caption{Working principle of Fourier-Domain OCT and its adaptation using a nonlinear interferometer.}
\label{fig:OCT}
\end{figure}

Optical coherence tomography (OCT) is an interferometric technique for 3D-imaging. It has a host of applications in non-destructive testing as well as medical imaging, e.g. in ophthalmology. In its most common modality (Fourier-domain OCT), spectral phases on broadband light proportional to the depth of reflections from inside a sample can be read out after interference with a reference arm using fast grating spectrometer. Fourier-transforming the spectra then yields an axial depth-profile of the sample, which after x-y-scanning of the sample and/or of the probing beam can be used to reconstruct a full 3D-image.
Since nonlinear interferometers by definition measure interferometrically encoded information, they are naturally suited to implement OCT - both in its time-domain\cite{Valles2018,Paterova2018_OCT} and frequency-domain flavour\cite{sven-oct}. The working principle of the latter is shown in Fig.\ref{fig:OCT}.

As with imaging applications, the main advantage of OCT with undetected light  is that one can illuminate the sample with light of a wavelength for which the detectors are not suitable. For example, infrared illumination can be desirable in the case of highly scattering (but water-free) samples, because scattering is strongly suppressed for longer wavelengths enabling high penetration depths.
In the practicle implementation of OCT with nonlinear interferometers, one of the crucial goals often is to reach a high axial resolution $\Delta z$, which is directly related to the idler spectral bandwidth $\Delta \lambda$. For a Gaussian spectrum it is given by:
\begin{equation}
\Delta z = \frac{0.44 \lambda_{I}^2}{\Delta \lambda} 
\end{equation} which will be slightly modified for differently shaped spectra. Formulated in the frequency domain, as a rule of thumb, one can estimate that a $10 THz$ spectral bandwidth corresponds to about $20\mu m$ axial resolution.
Typical non-degenerate SPDC sources, however, feature bandwidths on the order of $1 THz$, resulting in an axial resolution $\Delta z>100\mu m$, which is not sufficient for most real-world applications. Notwithstanding, there are several strategies to engineer sources with much broader spectra: 
\begin{itemize}
\item Choose crystals with a short length, L, because the bandwidth normally scales as $1/L$. Note, though, that this severely reduces the total brightness, which scales as $L^2$. Thus a 10-fold improvement in axial resolution would result in a 100-fold reduction in brightness making this a sub-optimal design strategy.
\item Choose chirped poling periods, which can lead to drastically increased bandwidth. Here, again, the penalty to pay is in spectral brightness compared to a crystal of the same length. Nevertheless, this reduction in brightness is is not as severe as for short crystals, making this a possible option, especially for reaching ultra-broad bandwidths.
\item Use signal-idler group velocity matched phase-matching\cite{Vanselow2019}, which yields ultra-broadband spectra without sacrificing spectral brightness. Also, the bandwidth and total brightness is still dependent on the crystal length but the trade-off is different here: as a special trait of this type of phase-matching, the bandwidth scales only with $1/\sqrt{L}$ , whereas the spectral brightness still scales with $L^2$ and therefor the total brightness with $L^{3/2}$. Thus, trading off a factor of 2 in axial resolution (with a 4x longer crystal) results in 8x more photons per pump power.
\end{itemize}
Other relevant parameters for implementing FD-OCT with nonlinear interferometers are the sensitivity (SNR), imaging depth, SNR-roll-off and speed. They are highly analogous to conventional FD-OCT, for which there is extensive introductory and overview literature, e.g. \cite{fujimoto2008}.

A special feature of FD-OCT with a undetected photons is: SPDC has no spectral fluctuations beyond the shot noise, as it is induced by the vacuum with a temporally constant spectrum. Moreover, any intensity noise of the pump laser  does not reduce the sensitivity, because it only affects the absolute intensity of the whole spectrum. On the other hand, frequency noise of the pump laser would need to be as large as the spectrometer resolution to have an effect on the sensitivity. Thus, the shot-noise-limit for the sensitivity of FD-OCT can be reached quite straight-forwardly\cite{Vanselow2019}. Note that by harnessing the high-gain regime of SPDC it is even possible to surpass the shot-noise limit in OCT \cite{Machado2020}.

For the practical implementation of FD-OCT with or without a nonlinear interferometer, the large bandwidths used mandate careful dispersion management. This can be done numerically, but in practice it is better to physically compensate as much dispersion as possible directly in the setup. Interestingly and sometimes usefully,  the dispersion in nonlinear interferometers can be compensated both in the idler arm, and/or in the signal arm. Note, that due to phase-mismatch away from the center of the SPDC bandwidth there is an additional dispersion term from the SPDC crystal to consider, in addition to the dispersion of the optical components (filters, lenses, dichroic mirrors) signal and idler photons pass through.

Another practical requirement for implementation is the use of a suitably narrowband pump laser: in order not to affect the imaging depth and sensitivity roll-off, it should be at least a factor of 2 narrower than the resolution of the spectrometer used to acquire the signal spectra.

\subsection{Spectral imaging}
One of the first applications of sensing with nonlinear interferometers has been the pioneering work of Kalashnikov et al.\cite{Kalashnikov:2016cl}. They used a highly multi-mode geometry, where a large, collimated laser pumps two relatively thin ppLN crystal in a gas-cell. In between the crystals the generated mid-IR light is partially absorbed by a pressure-controlled CO2-gas. Leveraging the spectral-correlations between the visible and mid-IR light and the correlations between angle and wavelength in the SPDC process, the absorption spectrum of CO2 at 4.3$\mu m$ as well as wavelength dependent refractive index could be deduced from the non-trivial circular interference fringes imaged on a CCD camera. A sensitivity of around $10^{-5}$ for measuring (n-1) and $0.1cm^{-1}$for the absorption coefficient could be reached with this method.

Extending imaging to spectral imaging faces the "challenge of dimensionality" of the final sensor. As imaging information is typically 2D and spectral data 1D, a 3D sensor array would be required to achieve a single-shot wide-field spectral imaging instrument. Such sensors unfortunately do not exist, but there are multiple possibilities to address this challenge: The most obvious one is a scanning approach, where a single-mode nonlinear interferometer is used the measure the spectral properties and the focused idler beam is scanned over the sample for spatial information. This approach is relatively simple and robust and scanning can be fast and is routinely used in scanning confocal/fluorescence microscopes. Another approach is to sequentially take wide-field images at different wavelengths either by spectral filtering\cite{sven-microscopy} or tuning the emission wavelength of SPDC e.g. by temperature tuning the crystal\cite{paterova2020hyperspectral}. In principle, other approaches are also possible and potentially useful, depending on the application. These include Fourier-domain spectroscopy imaging approaches or hybrid schemes (for example, one spatial and one spectral dimension on a CMOS camera and scanning one remaining spatial dimension by moving the sample.)

\section{Designing a setup for a specific application}\label{sec:design}

As discussed in Sec.\ref{sec:rig-theory}, sensing based on induced coherence without induced emission relies on particle correlations. These (classical and quantum) correlations depend on the physical properties of the experimental setup. In the case of SPDC, for example,  the bandwidth of the pump laser together with the length of the nonlinear crystal determine the spectral entanglement of the twin photons. Similarly the transverse spatial properties of the pump beam together with the dimensions of the nonlinear crystal determine the spatial entanglement. We emphasize, however, that spectral and spatial entanglement are not necessary for imaging with undetected photons. Classical correlations are enough.

\subsection{Wavelength considerations} 

Before getting started, the frequency range for probing the sample and for the detection need to be chosen, while also considering the resulting pump frequency given by energy conservation. It useful to consider a number of trade-offs and boundary conditions for this multi-parameter selection.

For the selection of the probing wavelength a range of special interest is the mid-IR region. Here, ro-vibrational molecular absorption lines, which are both strong and very specific for the type of molecule that is probed, are excellently suited for (bio-)chemical analysis by spectroscopic measurements. As a consequence of the transparency range of commonly used nonlinear crystals like ppKTP, ppLN or ppSLT, one can access wavelengths up to $\sim5\mu m$ , which contains important wavelengths used for gas sensing as well as the CH-stretch region around 3.45$\mu m$, which is useful for the identification and analysis of organic compounds, including (micro)plastics, lipids in tissue or collagen. Using nonlinear IR crystals such as AGS or opGAP, longer wavelengths in the so called "fingerprint region" (up to 12$\mu m$) can be generated and used.

Longer wavelengths scatter much less than shorter wavelength. This is useful, for example, for OCT to perform 3D-imaging into otherwise strongly scattering media like ceramics\cite{sven-oct}. Here the specific wavelength is less important than the bandwidth, which defines the axial resolution.
Depending on the specific application a large bandwidth is also relevant for spectroscopy or spectral imaging. Large bandwidths will be achieved if one matches the probing and detection wavelength's group indices in the nonlinear crystal\cite{Vanselow2019}. Interestingly, this can typically be achieved in  ppKTP, ppLN and ppSLT for a sensing wavelength in the mid-IR group-velocity matched with a wavelength in the NIR or SWIR \cite{Vanselow2019}. This strategy can also be applied for IR crystals like AGS (via angle phase-matching) or opGAP. This allows a certain design freedom to also accommodate other criteria in the wavelength choices, such as specific pump wavelength, water-absorption windows or the availability or cost of optical elements necessary.

Another possibly interesting choice for the sensing wavelength is the UV. This can increase the maximal resolution by leveraging the lower Abbe-limit for UV-Photons, while detecting in the technologically easier VIS/NIR region (see secs.\ref{subsec:momcorr-img-reso} and \ref{subsec:poscorr-img-reso}). Also, spectral imaging in the UV has applications in bio-imaging. The challenge here is the generation of photon pairs. In SPDC the pump wavelength needs to be shorter than the sensing wavelength, which has limitations due to the  transparency ranges for dielectric nonlinear crystals. A possible route around this is to use Spontanteous Four-Wave Mixing in gas-filled hollow-core fibers\cite{Lopez_Huidobro_2021}.

For the optimal choice of detection wavelength using Si-based CCD- and CMOS-sensors and cameras the detection wavelength should be chosen below  $\sim 900$nm. Above this wavelength the quantum efficiency of Si-based sensors typically drops quickly and vanishes above the Si bandgap ($\sim 1100$nm.) 

\subsection{Pump lasers}
Having selected the probing and detection wavelength automatically, the pump-wavelength is automatically determined via energy conservation. Selecting a suitable pump laser is highly important, with availability, maximum power, coherence length, beam-quality and cost as the main criteria for selection. Especially when using spatial entanglement, the pump beam needs to be of high quality. For spectroscopic application, the laser bandwidth (associated with the coherence length) needs to be below the target spectral resolution one wants to achieve. The same is true (in the time-domain) for OCT applications. 

\subsubsection{Spatial properties}
Perfect momentum correlation would ideally be obtained for a plane wave pump, which obviously cannot be generated perfectly experimentally. %Of course, lasers do not generate plane waves but beams that can be approximately described by paraxial gaussian beam formulas. 
However, this fact is an excellent guideline to what spatial properties of the pump beam are important for imaging applications: Ideally the pump beam is as spatially coherent as possible, its wave-front flat and with an only slowly varying amplitude. 

Lasers differ strongly in their beam quality depending on their type. Typical gas, solid state, and fiber lasers have very good beam quality, whereas most diode lasers (popular for their compactness and efficiency) have a far less ideal  beam quality. This can be remedied, if the pump laser we plan to use is filtered spatially before pumping the nonlinear interferometer. We can either use a 4f system with a pinhole in the Fourier plane or couple the laser beam into a single mode fiber. If the filter parameters are optimally matched to the incoming beam, only the higher order modes will be removed and the fundamental mode will be transmitted. Clearly, the losses in both methods will depend on the initial beam quality. One should take into account those losses when estimating the required laser power. 

\subsubsection{Spectral properties}
The spectral properties of the pump laser are essential for spectroscopy, OCT and spectral imaging, but it should be kept in mind that the coherence length of the pump can affect also the applications which are not directly related to spectral measurements (see Sec.\ref{sec:tips}\ref{sub:temporal-alignment}-\ref{sub:spatial-alignment}). Interestingly, because of the typical spectral-angular correlations in SPDC from a bulk crystal the spectral properties of the pump laser can affect the spatial properties of the SPDC. Therefore it is advisable to choose laser with  coherence length as large as possible, whenever available, and carefully consider the effects of using lasers with a shorter coherence length.
Another important feature can be wavelength stability, which if not sufficient can influence spectral/axial calibration of the spectroscopy/OCT data.

\subsection{Transverse spatial correlations}
The crucial element in the design of the interferometer for a given application is to provide correlations in the degree of freedom of interest. Spatial entanglement is in general not necessary for the applications of induced coherence without induced emission that we discuss in this tutorial (please see the explanation in Sec.\ref{subsec:gen-q-state} and Fig.\ref{fig:entanglement-explanation}). As shown in Secs.\ref{subsec:momcorr-img-reso} and \ref{subsec:poscorr-img-reso}, the resolution of the image is directly affected by the spatial correlations at the plane in which the sample is placed with respect to the crystal(s). 
%For good spatial entanglement the pump beam needs to be essentially single mode Gaussian ($TM_{00}$). 

%number of resolution elements (connect to Helen’s plot fig. 4; scanning vs widefield) - pump Width at crystal

The transverse momentum correlations in SPDC can be thought of as a consequence of the transverse momentum conservation in this process\cite{walborn10}. The sum of momenta of signal and idler photons equals the momentum of the pump photon. As a consequence, if the nonlinear crystal was pumped with a plane wave, which has a well-defined momentum, we’d obtain the perfect momentum correlation of the photon pair. Knowing the momentum of the signal photon, we could predict the momentum of the idler photon. Physically realizable beams have a finite spatial extent and as a consequence, their momentum distribution has to have a finite width. This uncertainty of the momentum of the pump beam translates to imperfect correlations of signal and idler beams. Knowing the momentum of the signal photon, we can predict the range of idler photon momenta; the correlation is not perfect. The better the pump beam can be approximated by a plane wave, the sharper the momentum correlation between the idler and signal photons.

In SPDC transverse position correlations are quite sensitive to the crystal length\cite{walborn10}. The shorter the crystal length, the sharper the transverse position correlations between signal and idler.  
\subsection{Nonlinear crystals}
There is a wide range of nonlinear crystals that have been used for SPDC and therefore are in principle suited for nonlinear interferometers, such as ppKTP, ppLN, ppSLT, BBO, BiBO, AGS just to name a few. However, when choosing a crystal optimally suited for the targeted application there are a number of important criteria to consider: 
\begin{itemize}
\item Achievable phase-matching - without phase-matching either by angle/bi-refringent or quasi-phasematching, no SPDC can be generated efficiently (except in ultra-thin crystals).
\item Suitable transparency range - the crystal should be highly transparent for the used wavelengths.
\item Brightness (nonlinear coefficient) - high brightness (i.e. a large photon-rate) will lead to desirable short measurement times.
\item Large bandwidth - the bandwidth determines both spectral range in spectroscopy applications as well as axial resolution in OCT and should therefor be as large as possible for those.
\item Crystal length and aperture - the maximum length co-determines the brightness trading-off with the bandwidth; the size aperture defines (together with the selected wavelengths) the number of spatial modes that for a certain length crystal can be generated and used for imaging\cite{kviatkovsky2021}.
\item General commercial availability and cost - important criterion for real-world applications. Typically, however, pump laser and final sensor will have a more poignant influence on the cost.
\end{itemize}

\subsection{Cameras}
Typically,  Si-based detectors are used for detecting the signal photons, as the detection wavelength is often purposefully chosen to lie in their sensitivity range. If cameras are used, sCMOS cameras are often preferable to the EMCCD cameras, because typically a photon flux per pixel of more then 10-100photons/s can be easily achieved, and sCMOS cameras feature a better SNR at such illumination levels than EMCCDs and can have more and smaller pixels. In general, the number of illuminated pixels (magnification onto the camera) should be chosen such that 1 spatially resolved element corresponds to 2-3 Pixels. Having more pixels per resolved element, does not bring any benefit, but on the contrary increases the amount of total readout and dark noise from the camera and therefore lowers the SNR.

\section{Experimental guidelines: tips and tricks}\label{sec:tips}
Most of the tips and tricks we provide below for the ZWM mandel setup, also apply to the SU(1,1) interferometer in the low gain regime, such as those illustrated in Figs.\ref{fig:herzog}, \ref{fig:OCT}b  and \ref{fig:herzogalternative}. Notice that in the setup in Fig.\ref{fig:herzogalternative}b all three fields, signal, idler and pump, go through the imaged object, which will  lead to further effects not described in this tutorial.

\subsection{Building and aligning a Zou-Wang-Mandel interferometer}
Setting up a Zou-Wang-Mandel interferometer can be challenging because the alignment needs to be performed with SPDC light, not with bright laser beams. Moreover, correlations required for the multimode sensing applications can result in a low degree of spatial and spectral coherence of the individual photon beams. As a consequence, the visibility of interference depends critically on transverse and longitudinal shifts between the interfering beams\cite{barbosa1993degree,grayson1994spatial}. In this section, we give some tips and tricks for building and aligning a Zou-Wang-Mandel Interferometer,  such as those in Figs.\ref{fig:basicmandel} and \ref{fig:img-schm-mc}.

\subsubsection{Controlling the relative phase and power between pair emissions at the two sources}

In principle, a $50:50$ beam splitter could be used to split the pump beam in the ZWM interferometer and the phase between the interferometric arms could be scanned by shifting this beam splitter or one of the mirrors, as shown in Fig.\ref{fig:basicmandel}. In practice, however, it is advantageous to have fine-tuned control of the relative pump power illuminating the crystals. In order to do this, we recommend using a half-waveplate ($HWP_1$) followed by a polarizing beam splitter ($PBS$) instead of $BS_1$, as shown in Fig.\ref{fig:img-schm-mc}. Rotating $HWP_1$ not only enables finding the optimal ratio of powers for the maximal visibility, but also can be used to aid alignment, by directing all the pump power to one of the crystals. This can be helpful for example when overlapping the SPDC beams from both crystals. Fig.\ref{fig:img-schm-mc} also shows another half-waveplate ($HWP_2$) at $45^{\circ}$ placed in one of the pump paths, which guarantees that both crystals are pumped with the same polarization. 

It is also useful to be able to shift the interferometric phase without touching any mirrors or beamsplitters. For that, one can use a single quarter-waveplate ($QWP$) set at zero degrees before the PBS and rotated around its vertical axis. Alternatively, one can place before the PBS a combination of another half-waveplate sandwiched between two $QWP$s at $45^{\circ}$. The interferometric phase is then tuned by rotating the sandwiched $HWP$.

\subsection{Temporal alignment}
\label{sub:temporal-alignment}
\subsubsection{From indistinguishability to zero-delay requirements}

In building any interferometer, it is necessary to adjust the combined path lengths to within the coherence length of the detected light in order to observe interference. In the language of quantum information, this condition is due to the requirement of path indistinguishability \cite{englert1996fringe}. Let us consider the ZWM interferometer shown in Fig.\ref{fig:basicmandel}, and denote $l_{P_1}$ and $l_{P_2}$ as the optical path lengths of the pump from $BS_1$ to nonlinear sources $Q_1$ and $Q_2$, respectively. We will denote $l_{I_1}$ as the idler optical path in between the two sources, and we shall label the signal $S_1$ ($S_2$) optical path length from $Q_1$ ($Q_2$) to $BS_2$ as $l_{S_1}$ ($l_{S_2}$). Two path length requirements can be formulated \cite{wangthesis}: 
\begin{eqnarray}
\delta l_1=l_{P_1}+l_{I_1}-l_{P_2} < l_{coh-pump};\label{eq:lengthcond1}\\
\delta l_2=l_{I_1}+l_{S_2}-l_{S_1}<l_{coh-signal}, \label{eq:lengthcond2}
\end{eqnarray}
 where $l_{coh-pump}$ and $l_{coh-signal}$ are the pump and signal field coherence lengths, respectively. 

These conditions can be understood intuitively, by analyzing how the two interfering alternatives (emissions from the two crystals) could be distinguished.To find the first condition (Eq.\ref{eq:lengthcond1}), we consider a pulsed excitation of the two crystals. We require the idler photon generated in the first crystal to arrive at the second crystal together with the pump pulse, so that the idler photons generated in the two crystals overlap in time. Strictly speaking, it is the coherence length of the pump laser, not the pulse length that governs the second condition.  Note that condition (Eq.\ref{eq:lengthcond1}) is valid for the case of  excitation using  so-called continuous wave (CW) lasers.
%I want to reformulate this without pulses.
To find the second condition, (Eq.\ref{eq:lengthcond2}) let’s consider a single pair of photons generated in one of the crystals. If we were to detect both photons, we \textit{could} measure the time difference between their arrival at the detectors and, if this time was different for the two pairs, we could distinguish between the two alternatives. Note, however, that even if we do not detect idler photons, just the possibility of obtaining \textit{welcher weg} information is enough to inhibit interference in the signal. 

We stress again that it is not relevant in this reasoning if we actually perform the measurements required to distinguish the alternatives or not. The mere possibility of performing such measurements precludes the observation of interference or in the intermediate case reduces the visibility of interference fringes. For a detailed theoretical treatment please refer to \cite{Wang:1991p625, wangthesis,zouthesis}.

\subsubsection{Finding the zero delay using the fringe visibility or spectral interference}

In order to adjust path lengths for the conditions expressed above, one can maximize the visibility of interference fringes as a function of an optical delay added in the signal, idler and/or pump modes, or, alternatively, by observing the spectrum of the interfering signal beams from the two crystals. We recommend this latter strategy if interference is not immediately seen in the detector/camera.  The non-zero path length difference $\delta l$ (corresponding to non-zero optical delay difference $\tau$ in the interferometer) results in the appearance of spectral interference fringes with the period given by $\frac{2 \pi}{\tau}$. The effect of the spectrometer is as if one had many bandpass filters, one for each wavelength to within the spectral resolution of the equipment. In other words, the effective coherence length of the fields in the equations above is larger than that which most bandpass filters can provide. Therefore, one sees interference in the spectrometer before one can observe interference in the detector. If the resolution of the spectrometer is high enough to resolve the fringes for a given delay, it is straightforward to minimize the optical path length difference, by maximizing the period of spectral interference fringes up to a point when the entire spectrum is in phase. In other words, as the interferometer path lengths are adjusted, the interference fringes become spectrally wider in the spectrometer. In fact, their width is inversely proportional to the path length difference $\delta l$.   Of course, the broader spectrum results in shorter coherence length and in the requirements for the observation of interference becoming more stringent.

\subsubsection{Spectral bandpass filters for the initial alignment}

Because SPDC in general produces broadband fields, it is useful to use narrow bandpass filters at the detectors in order to have a workable signal coherence length. The narrower the interference filter in front of the camera the easier the alignment is. Nevertheless, the final alignment needs to be performed with a spectral bandwidth broad enough for the final application.

\subsection{Spatial alignment}
\label{sub:spatial-alignment}
\subsubsection{Alignment of the imaging systems}

Imaging requires multiple imaging and/or Fourier transform optical systems (Fig.\ref{fig:img-schm-mc}). Each of these lenses or off-axis parabolic mirror systems has to be aligned separately. In the case of perfectly aligned imaging systems in the interferometer we should obtain an unmodulated intensity profile at the camera. Misalignments in the setup can in turn lead to the appearance of intensity fringes at the camera plane, which for the case of the momentum correlation enabled QIUP setup were studied in \cite{Hochrainer:17}. In the ZWM interferometer, misalignment of the signal field can, to a certain extent, be ''compensated'' with the misalignment of the idler field. This is quite unfortunate because one can end up with sub-optimal interference visibility. 

Typically the dichroic mirrors and beam splitters allow one to perform the initial alignment of the setup using the pump beams. In this case, the alignment is not different than that of the standard Mach-Zehnder interferometer. Later the idler beams need to be overlapped to ensure the indistinguishability of the idler photons generated in the two crystals. It is convenient to overlap the two beams by monitoring them in two complementary bases, for example at an image plane of the crystal and at a Fourier plane of the crystals.  For this alignment it is very helpful to have two lenses on flipper mounts just in front of the camera, one that images onto the camera the crystal plane and the other performing the Fourier transform of the electric field at the crystal plane. 

In certain designs (see for example \cite{kviatkovsky2021}), beams with very different wavelengths pass through certain imaging components. In such cases chromatic aberrations need to be minimized by using achromatic lenses or replacing the lenses with off-axis parabolic mirrors.

\subsubsection{Coupling between spatial and temporal alignment by thick lenses}

We often describe optical setups using thin lens approximation. In practice, when we want to minimize the distances in the setup, we use lenses with short focal lengths. Such lenses, because of their thickness, introduce a significant delay that depends on the distance from the optical axis. This can lead to complications when aligning the setup if one is not careful to always keep the optical axis of the propagating fields aligned with the center of the lenses. %adjusting the optical path lengths accordingly by any means. 

Standard lens alignment procedures are helpful here: preparing a laser beam along the propagation direction of the beam of interest and making sure that the beam pointing direction is not modified when an extra lens is introduced and at the same time that the back-reflection from the lens surface comes back very close to the source. If back-reflections from both surfaces of a lens are visible, it suffices to overlap them.

\subsubsection{Spatial partial incoherence of the signal and idler beams}

In the case of spatially multimode SPDC, which is necessary for widefield imaging, individual signal and idler beams are spatially partially incoherent. This has profound implications for the design of the experimental setup, as partially incoherent light propagates differently than perfectly coherent light. In particular, one should expect a much larger illumination area on the object (FoV) than if these were propagating coherent Gaussian fields.  In other words multimode SPDC light doesn’t form a beam that can be treated as a laser beam, the latter being fully spatially coherent.

\subsubsection{Using an auxiliary laser beam to stimulate emission into idler and signal beams for easier alignment}

Parametric downconversion can be induced (stimulated) by aligning an auxiliary laser source with at least one of the downconverted beams generated in the crystals. The process of stimulated parametric downconversion is a useful tool for the alignment of the ZWM or SU(1,1) interferometer (Fig.\ref{fig:entanglement-explanation}b) because it significantly increases the intensity of the detected signal and allows one to increase the coherence length of the PDC light. Moreover, if the stimulating laser has a larger coherence length than the band pass filter used in front of the camera, the resulting longer coherence length of the stimulated beams allows the experimentalist to focus on the spatial part of alignment before dealing with the fine adjustment of the 
optical path lengths (temporal alignment).

Stimulated emission inherits also the spatial coherence properties of the stimulating beam, therefore the stimulation with coherent laser beams results in a down-converted beam with a narrower angular spread than in the case of the spontaneously emitted light. When spots from the two crystals are being overlapped at the camera, the smaller spread translates to a possibility of a more precise alignment. Therefore it is particularly useful to combine the stimulated emission with monitoring the Fourier and image planes.

The "trick" of using a seed laser can also be used for imaging via induced coherence \textit{with} induced emission, as was demonstrated in ref.\cite{Cardoso2017}. Their setup is illustrated in Fig.\ref{fig:entanglement-explanation}(b). The sample is illuminated with the seeding laser. Detection, pumping and seeding wavelengths are chosen according to one's needs, taking into consideration the availability of lasers, crystals and detectors. Notice that cheaper cameras can be used, as the signal is much brighter than in the case of spontaneous down-conversion. 

\subsection{Loss within the interferometer}

In a standard interferometer, loss in one of the arms can be compensated for by introducing an equal loss in the other arm. It is a distinguishing feature of the ZWMI and other interferometers based on induced coherence without induced emission, that losses in the idler arm between the two crystals lead to a reduction in the mutual coherence of the two signal beams and therefore cannot be compensated. As a consequence, it is essential to minimize the losses in the path between the two crystals (or equivalent paths in the designs different than ZWM), as the transmission of this arm sets the upper bound for interference fringe visibility.

To sum up, in the ZWM experiment, losses in the idler path between the crystal (refer to Fig. 1) control the mutual coherence between the signal beams and cannot be compensated for, whereas the losses in other arms do not affect the coherence and can be compensated.

\section{Outlook}\label{outlook}

One of the most important ways to categorize various types of interferometers (both linear and nonlinear) is how they measure up against specific limitations, and whether or not they exceed them. Based on what application is in mind for a device it can be measured up against one, or more, of several types of limits on the ability of the device to perform specific tasks \--- typically related to the sensitivity, signal-to-noise ratio,  and/or resolution.

With this in mind, there are a number of interesting directions where the field is headed. Firstly it will have a major impact when experimental techniques improve sufficiently to robustly enter the high-gain regime. Sub-shot-noise performances in different sensing arrangements, as well as compensating SNR reductions due to losses and in general much higher photon-rates, leading to shorter measurement times, will be in  focus then. Moreover, extending the wavelengths that are covered into the finger-print region beyond 6$\mu m$ as well as further improving techniques in the THz region.

Moreover, a number of further interferometric sensing and imaging techniques could be adapted for nonlinear interferometers, such as tomographic holography for 3D imaging, a combination of coherence gating (with a broadband source) and holography, to achieve optical sectioning as well as aperture synthesis techniques like Fourier ptychography.
Finally combining nonlinear interferometry with super-resolution techniques might lead to novel imaging modalities to be explored.
\par
In addition to applications to imaging and metrology, the study of ZWM-type nonlinear interferometers have had important implications to fundamental quantum physics. The ZWM-interferometer has already been used to study quantum complementarity and its connection to spatial coherence \cite{mandel1991coherence,zou1991} and partial polarization \cite{lahiri2011wave,Lahiri2015b}. Similar setups have been recently used to measure twin-photon correlations \cite{Lahiri2016,Armin2016a} and entanglement \cite{lahiri2021characterizing,lemos2020measuring}. Generalizations of the ZWM-experiment have led to the discovery of novel methods of generating multi-particle and/or high-dimensional entangled states \cite{Mario2016,Lahiri2018,kysela2020path}. Therefore, further studies of such nonlinear interferometers will potentially result in new research direction in fundamental quantum physics and quantum information science. 
\begin{backmatter}

\bmsection{Acknowledgments} We thank  Inna Kviatkovsky, Helen M Chrzanowski, Victoria Borish, Jorge Fuenzalida, Armin Hochrainer, and Anton Zeilinger for numerous discussions. G.B.L. acknowledges support from the Brazilian National Council for Scientific and Technological Development (CNPq),  from the Fundação de Amparo à Pesquisa do Estado do Rio de Janeiro - FAPERJ, and from the Coordenação de
Aperfeiçoamento de Pessoal de Nível Superior (CAPES - Brasil) -- Finance Code 001. R.L. acknowledges the funding by the Foundation for Polish Science within the FIRST TEAM project `Spatiotemporal photon correlation measurements for quantum metrology and super-resolution microscopy' co-financed by the European Union under the European Regional Development Fund (POIR.04.04.00-00-3004/17-00).
S.R. acknowledges funding from Deutsche Forschungsgemeinschaft (RA 2842/1-1) as well as from Bundesministerium fuer Bildung und Forschung (BMBF) within projects QUIN (13N15402) and SimQPla (13N15944).
\bmsection{Disclosures} The authors declare no conflicts of interest.

\end{backmatter}

\bigskip

% Bibliography
\bibliography{tutorial_imaging.bib}

% Full bibliography added automatically for Optics Letters submissions; the following line will simply be ignored if submitting to other journals.
% Note that this extra page will not count against page length
\bibliographyfullrefs{sample}

\end{document}